\documentclass[fleqn,10pt]{wlscirep}
\usepackage[utf8]{inputenc}
\usepackage[T1]{fontenc}
\usepackage[labelformat=empty, position=top]{subcaption}
\usepackage[export]{adjustbox}
\usepackage{lineno}
\usepackage{cleveref}

\title{Product-level value chains from firm data: mapping trophic levels into economic growth}

\author[1]{Massimiliano Fessina}
\author[2]{Andrea Tacchella\thanks{Corresponding author: \texttt{andrea.tacchella@cref.it}}}
\author[3]{Andrea Zaccaria}

\affil[1]{IMT School for Advanced Studies, 55100 Lucca (Italy)}
\affil[2]{Enrico Fermi Research Center (CREF), 00184 Rome (Italy)}
\affil[3]{Institute for Complex Systems (CNR-ISC), UOS Sapienza, 00185 Rome (Italy)}

\begin{abstract}

We reconstruct a product-level input-output network based on firm-level import-export data of Italian firms. We show that the network has a statistically significant, yet nuanced trophic structure, which is evident at the product level but is lost when the classification is coarse-grained. This detailed value chain allows us to characterize the trophic distance between inputs and outputs of single firms, and to derive a coherent picture at the sector level, finding that sectors such as weapons and vehicles are the ones with the largest increase in downstreamness between their inputs and their outputs. Our measure of downstreamness at the product level can be used to derive country-level indicators that characterize industrial strategies and capabilities and act as predictors of economic growth. With respect to the standard input/output analysis, we show that the fine-grained structure is qualitatively different from what can be observed using sector-level data. We finally prove that, even if we leverage exclusively data from Italian firms, the metrics that we derive are predictive at the country level and capture a significant description of the input-output relations of global value chains.
\end{abstract}

\begin{document}

\flushbottom

\maketitle
\thispagestyle{empty}


\section*{Introduction}
Supply chains represent a crucial element of the economy, both at local and global level, by ensuring a seamless flow of goods, services, and information between economic actors. A close concept is that of value chains, where the focus is centered on the value that is added at every stage of the production process, through the transformation of inputs into outputs. They constitute a key aspect of companies’ diversification and innovation strategies and are pivotal for both profitability and sustainability studies. Both systems, as suggested by their names, are commonly depicted as sequential structures, with a clear direction leading from raw materials, through increasingly complex goods, up to finite products, being sold to final customers. This picture raises the natural question of where industries place themselves along these chains, and what the economic implications of their position are.

Starting with the seminal work by Antras and Chor\cite{antras2012measuring}, that introduces a measure of \emph{upstreamness} of an industry, i.e. its distance from final demand, and relates it to the quality of institutions and skilled labor, several studies proposed metrics to assess the position of industries along value chains. Fally\cite{fally2012production} introduces the complementary measure of \emph{downstreamness}, i.e., the distance of an industry from basic goods, showing that upstream industrial stages contribute less to the final value, and how the most developed countries tend to specialize in downstream industries. Miller and Temurshoev\cite{miller2017output} distinguish between input (demand) and output (supply) chains, finding a positive correlation between the output downstreamness and the input upstreamness across industries. McNerney et al.\cite{mcnerney2022production}, building on the evidence that later (i.e. more downstream) industrial stages add more value to products, propose a model to relate the average length of a country's supply chain to its economic growth. However, Bartolucci et al.\cite{bartolucci2023correlation,bartolucci2025upstreamness} show how the aggregate properties of both upstreamness and downstreamness can be traced back to the minimal structural constraints satisfied by inter-industry monetary flows. All these investigations are based on Input/Output tables, a well-known database\cite{leontief1986input} that maps input-output relationships between industrial sectors at national\cite{antras2012measuring,fally2012production} and international\cite{miller2017output,mcnerney2022production,bartolucci2023correlation,bartolucci2025upstreamness} level. The main limitation of this database lies in its aggregation level: the number of industries is in the order of tens, while thousands of different products are exchanged - and properly classified in other contexts, such as global trade.

Indeed, scholars have recently pointed out the inadequacy of industry-level data for the study of supply chains\cite{acemoglu2012net}, triggering a flourishing literature on firm-level production networks, focusing on systemic risk\cite{inoue2019firm,diem2022quantifying,fessina2024inferring,inoue2023simulation,pichler2022forecasting}, network reconstruction\cite{ialongo2022reconstructing,mungo2024reconstructing,kosasih2021machine,wichmann2020extracting} and economic inequality\cite{chakraborty2024inequality}. However, despite the increasing availability of granular firm-level datasets \cite{bacilieri2022firm}, information on products is seldom available\cite{fessina2024pattern,brintrup2015nested}, forcing scholars to rely on the approximation that firms produce a single product, coinciding with its industrial classification, which has been shown to provide a distorted picture of actual input/output relationships\cite{diem2024estimating}. In conclusion, from the literature, a clear necessity for a more detailed view of the input-output relationship among single products emerges. In this respect, Karbevska and Hidalgo\cite{karbevska2025mapping} recently proposed a machine-learning-based reconstruction of a product-level value chain starting from international export data. However, this reconstruction suffers from the well-known limitations of country-level data, which presents spurious co-occurrences of products that are simply due to the heterogeneous diversification patterns of countries \cite{albora2022machine}.

This study aims to address this gap, leveraging a unique database that reports the import and export of practically all Italian firms, at the highest level of product detail (6-digit Harmonized System classification). By matching firms' imports and exports, we are able to construct a directed network of approximately 5000 different products: a link between two products is present if the number of firms importing the former and exporting the latter is statistically significant with respect to a suitable null model, rooted in the exponential random graphs framework\cite{squartini2011analytical,cimini2019statistical,saracco2015randomizing,saracco2017inferring,cimini2022meta}.

This association between products, de facto, reconstructs a \textit{book of components} for each product and the value chain for all manufacturing sectors. This network enables the computation of the production stage of each good through \textit{trophic levels}\cite{levine1980several,mackay2020directed} - a metric borrowed from the field of ecological networks that generalizes the concepts of upstreamness and downstreamness - allowing an assessment of the overall hierarchical organization of the value chain. The structure we find strongly depends on the products' aggregation level, pointing out the bias introduced by the widespread use of coarse-grained data. Trophic levels allow us to characterize the distance between inputs and outputs of firms through their \emph{trophic jump} - i.e. the difference between the average trophic level of their exports and their imports - which we find to vary coherently across industrial sectors. Moreover, we show that the trophic position of countries, measured through the average trophic level of their imports and their \emph{trophic jump}, is strongly related to their industrial development, representing a strong predictor of future economic growth.


\section*{Results}

\subsection*{Trophic levels in a product-level value chain}

The ISTAT \textit{Commercio con l'estero} dataset (www.coeweb.istat.it) details product-level information about practically all exporting and importing firms in Italy: we establish a link from one product to another if we detect a statistically significant number of firms that import the former and export the latter (see Methods). This way we construct a directed, input-output network of products encompassing roughly 5000 goods and 454.000 links between them, with an average degree $\langle k\rangle=91$. 
The computation of trophic levels (see \cref{eq:lapl,eq:troph_lev}) amounts at measuring the hierarchical position of each node if we try to arrange the network as a tree: products with the highest trophic levels are leaf nodes with no out-going links, while products with the smallest trophic levels correspond to root nodes with no incoming link (the minimum is set to 1, see Methods). In economic terms, the latter can be seen as \emph{basic goods}, acting only as input for other products; as for the former, they represent final products, intended for final consumption (see table \ref{tab:troph_lev}). Overall, trophic levels follow a right-skewed normal distribution, with mean value $\overline{\text{TL}}=2.83$ (see supplementary figure S1). 

\begin{figure}
    \centering
    \includegraphics[width=0.7\linewidth]{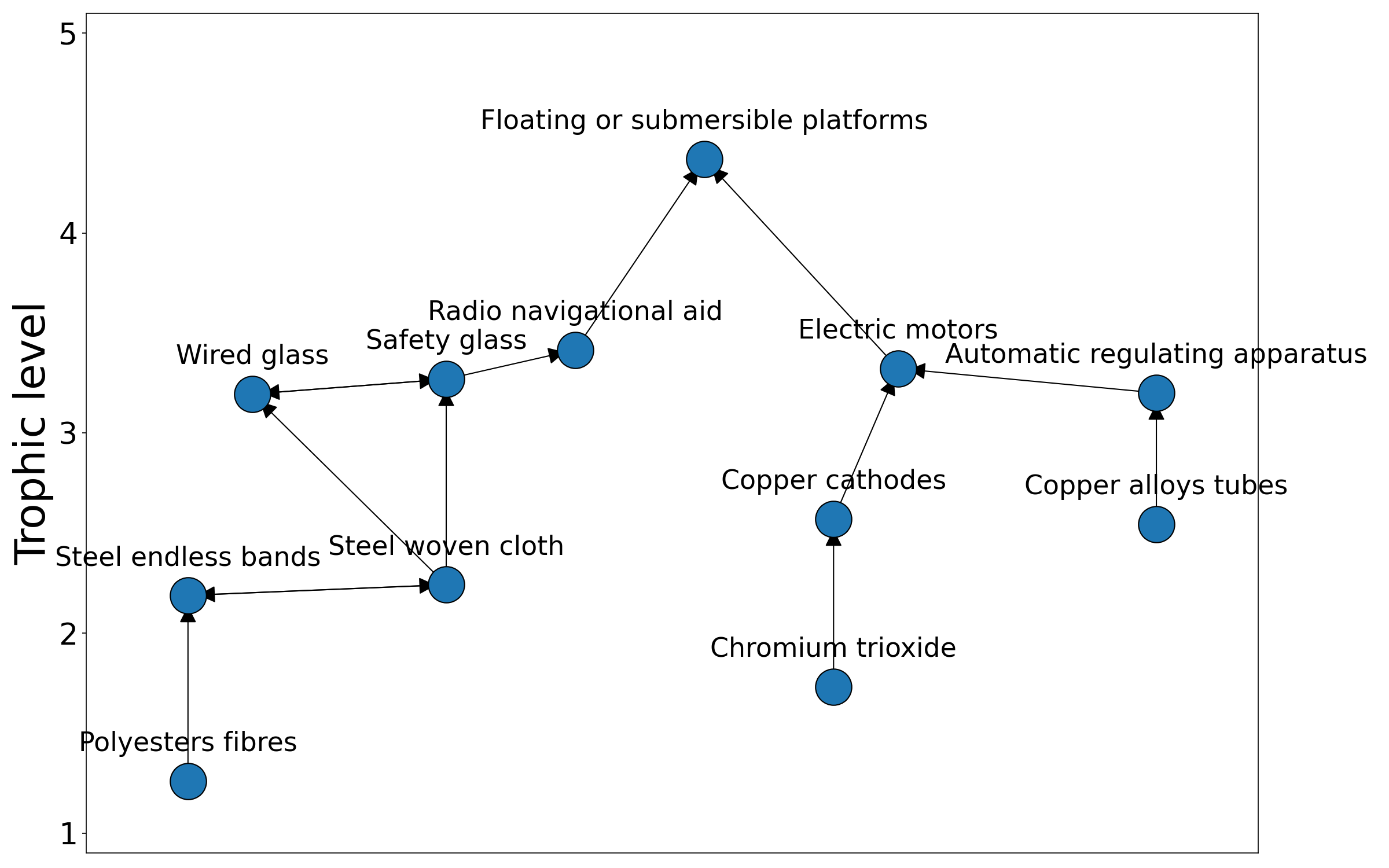}
    \caption{Small subgraph of the directed network of products we reconstruct. Here, a link is present if a statistically high number of firms import the source node and export the target node. In this layout, we order the nodes vertically using the trophic level. This induced hierarchy organized the products from basic inputs (at the bottom), to increasingly complex components up to products for final consumption (at the top).}
    \label{fig:trophic_levels}
\end{figure}

An example of the fine-grained hierarchy induced by trophic levels is displayed in figure \ref{fig:trophic_levels}, where we plot a small subgraph of the network: two root nodes, \emph{polyesters fibres} and \emph{chromium trioxide}, serve as input for metallic components (\emph{steel woven cloth}, \emph{copper cathodes}), in turn entering into glass components and complex machinery (\emph{safety glass}, \emph{radio navigational aid} and \emph{electric motors}), finally leading to the leaf node \emph{floating or submersible platforms}. The subgraph is obtained by choosing a final product (\emph{floating or submersible platforms}), moving backwards along two of the many possible links, and so on - as such, it shows only one of the many possible trees ending up in \emph{floating or submersible platforms}. Notice how a root node with a single link towards a product with a high trophic level will inherit a high trophic level itself: \emph{chromium trioxide} and \emph{polyesters fibres} are both root nodes, but with different trophic level.

\begin{table}[t!]
\centering
\begin{tabular}{lcc}
\hline
 Product & Trophic level & Ranking \\
\hline
950810 (\textbf{Traveling circuses}) & 4,84 & 1 \\
890520 (\textbf{Floating or submersible platforms}) & 4,37 & 2 \\
860390 (\textbf{Other van/coaches}) & 4,30 & 3 \\
... & & \\
550130 (\textbf{Acrylic filaments}) & 1,10 & 4.964 \\
530121 (\textbf{Scutched flax}) & 1,09 & 4.965 \\
510220 (\textbf{Coarse animal hair}) & 1,00  & 4.966 \\
\hline
\end{tabular}
\caption{Products with the highest and lowest trophic level: the former are final products, while the latter are basic inputs (interestingly, all belonging to the textile sector).}
\label{tab:troph_lev}
\end{table}

\begin{figure}
    \centering
    \includegraphics[width=0.7\linewidth]{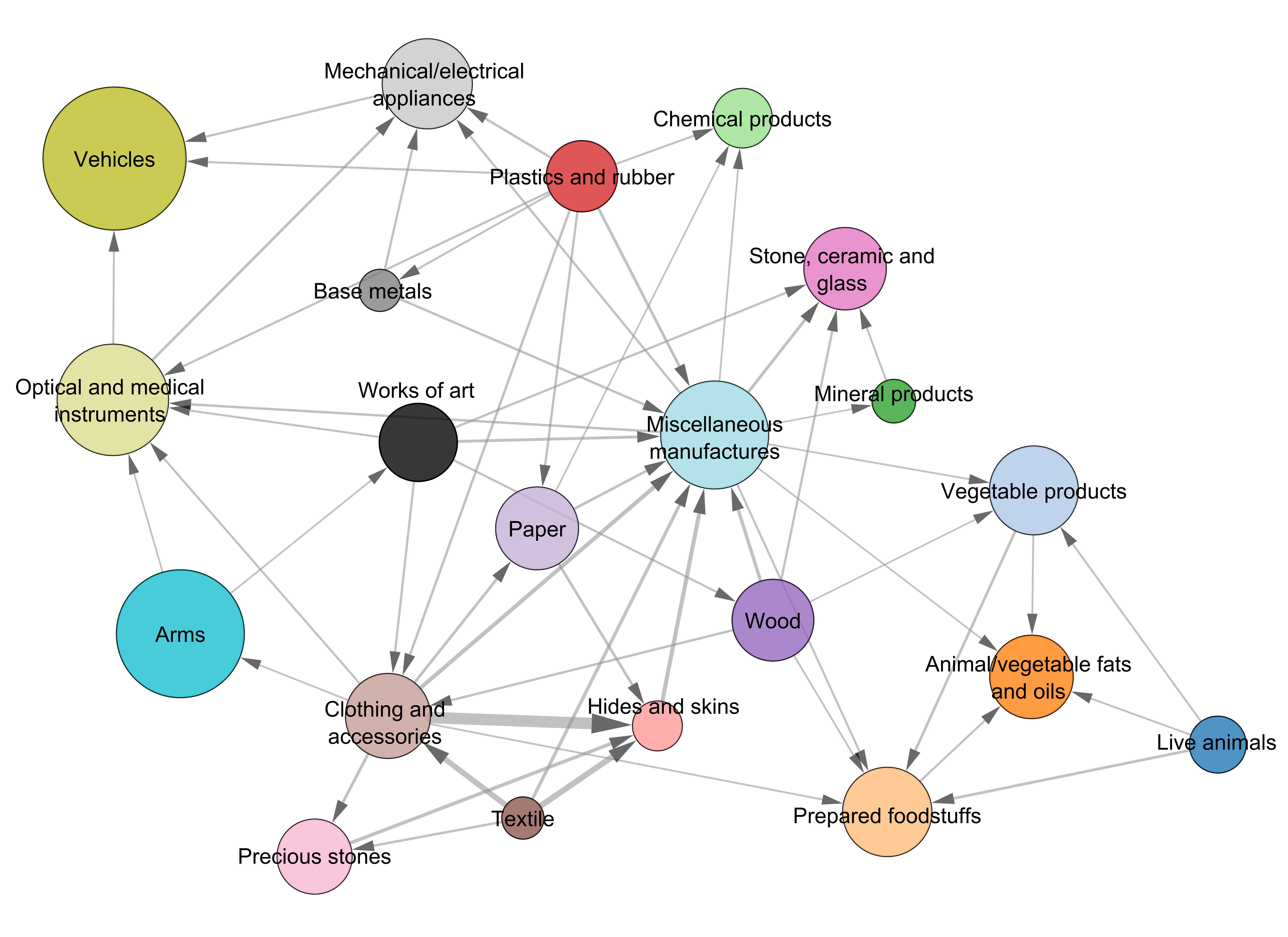}
    \caption{Products' network aggregated at the section level, filtered with the \emph{PMFG} method \cite{tumminello2005tool}: the size of nodes is proportional to the average trophic level of the products they encompass (the bigger the dot, the more downstream the sector), the width of the arrows to the intensity of links. }
    \label{fig:sect_net}
\end{figure}

In figure \ref{fig:sect_net} we display the network aggregated at the coarser sector level (called \emph{sections}, see supplementary section S1), filtered with the \emph{PMFG} method \cite{tumminello2005tool}. Although the graph does not display a clear trophic direction, the average trophic levels of sectors (displayed by the size of the nodes) reflect the distinction between primary industries (such as \emph{Base metals}, \emph{Mineral products}, \emph{Textile}), appearing mainly as \emph{sources} of links, and sectors closer to final consumption (such as \emph{Arms} and \emph{Vehicles}), in turn acting mainly as \emph{targets} of links. Furthermore, the majority of links mimic meaningful input-output dependencies, e.g. \emph{Live animals} $\rightarrow$ \emph{Animal fats and oils} and \emph{Prepared foodstuff}, \emph{Base metals} and \emph{Plastics and rubber} $\rightarrow$ \emph{Mechanical/electrical appliances} $\rightarrow$ \emph{Vehicles} and so on. It is interesting to notice the central role played by \emph{Miscellaneous manufactures}, a heterogeneous section grouping products such as lamps, prefabricated buildings, pens, etc., which receives input from many sectors and acts as input for many others. 

\subsection*{Trophic coherence}

In order to quantify to which extent the network exhibits an overall trophic structure, we measure its \textit{trophic coherence} \cite{mackay2020directed} (TC), computed as the average square distance between the difference among trophic levels of connected nodes and 1 (see \cref{eq:troph_coh}). A value of TC closer to 1 implies a well-defined hierarchical organization of the nodes into distinct levels - in other words, that the network exhibits a tree-like structure.

\begin{figure}[t!]
    \centering
    \begin{subfigure}{0.02\textwidth}
    \textbf{a)}
    \end{subfigure}
    \begin{subfigure}[t]{0.47\textwidth}  \includegraphics[width=\textwidth,valign=t]{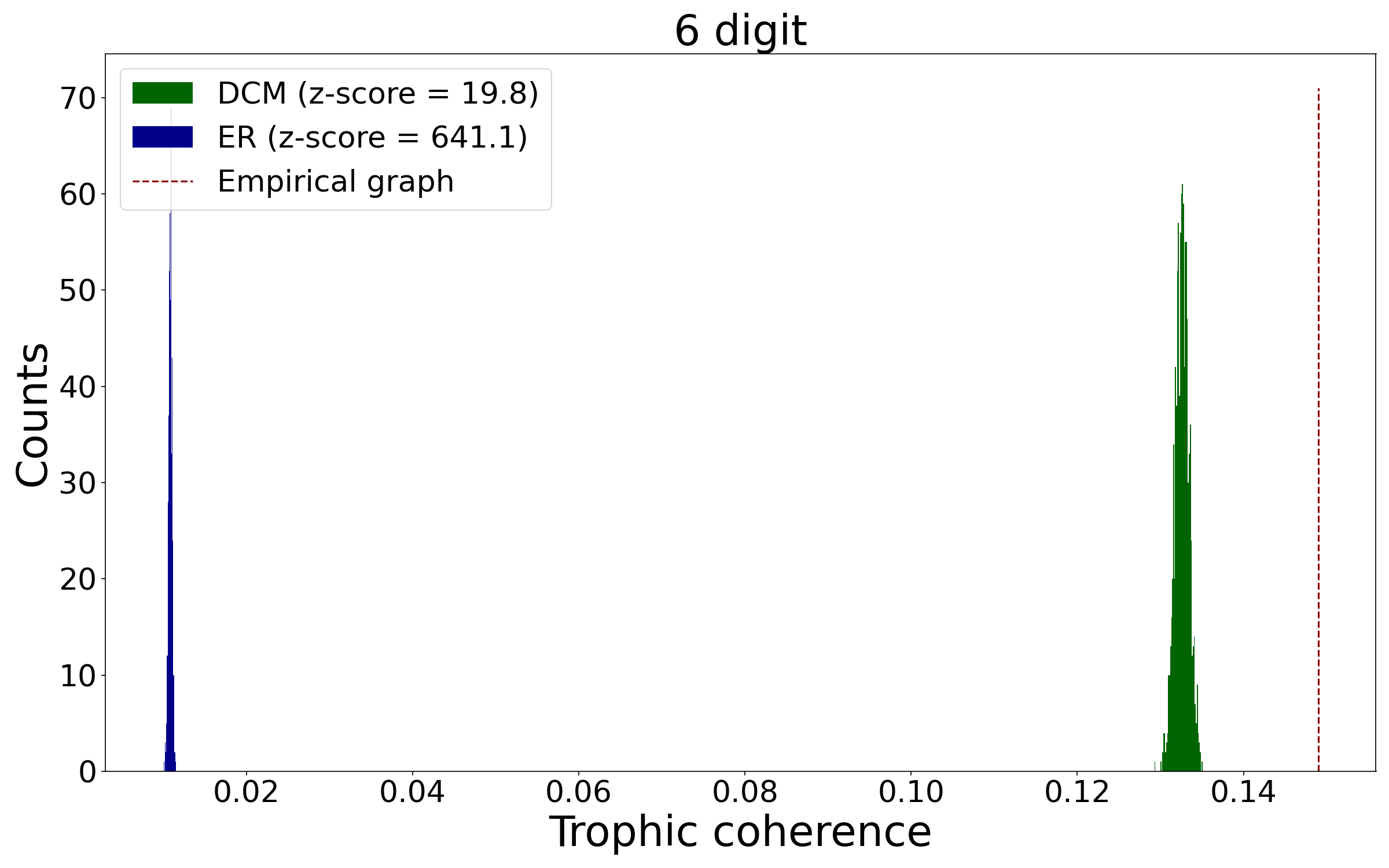}
    \end{subfigure}
    \begin{subfigure}{0.02\textwidth}
    \textbf{b)}
    \end{subfigure}
    \begin{subfigure}[t]{0.47\textwidth}  \includegraphics[width=\textwidth,valign=t]{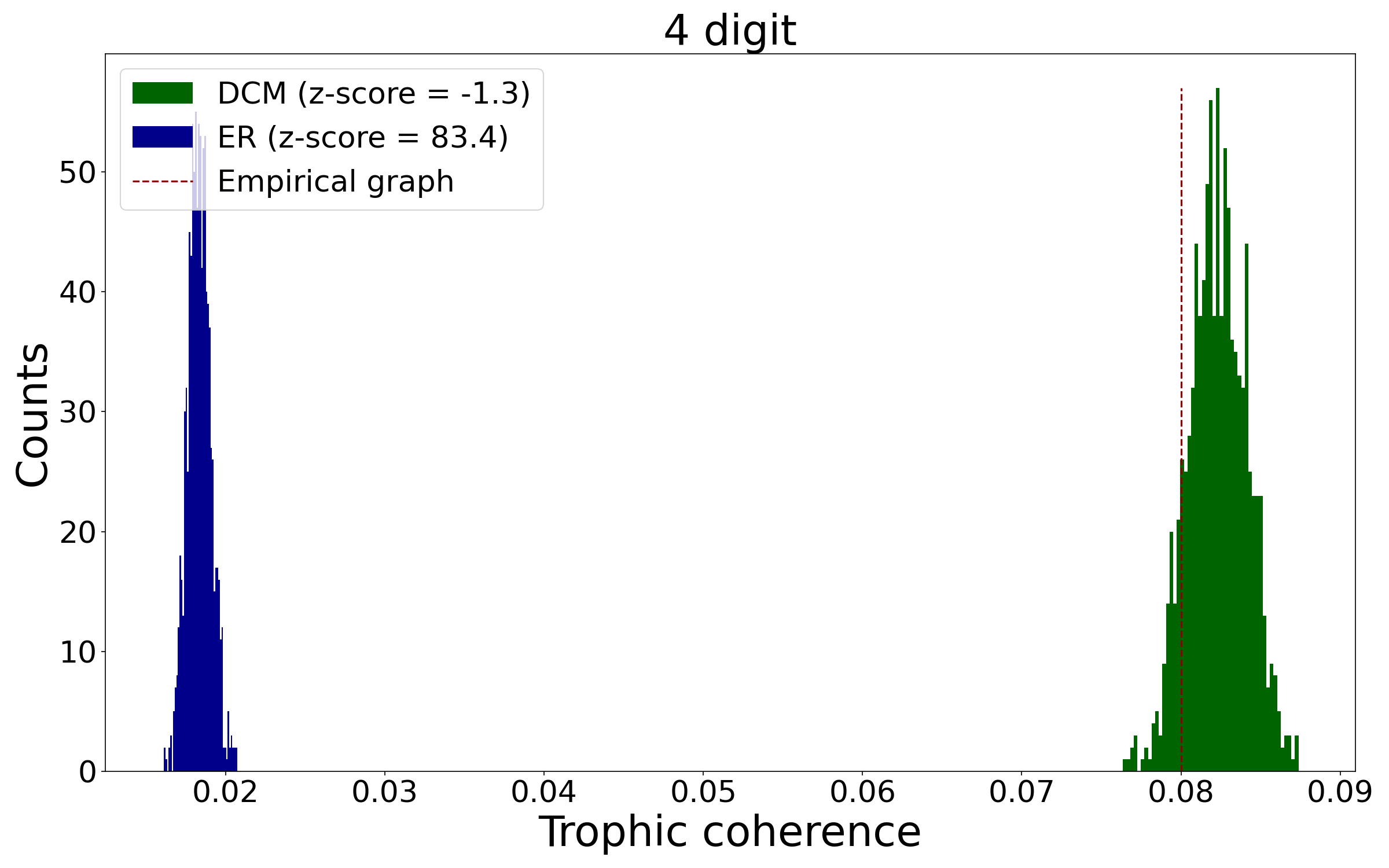}
    \end{subfigure}
    \begin{subfigure}{0.02\textwidth}
    \textbf{c)}
    \end{subfigure}
    \begin{subfigure}[t]{0.47\textwidth}  \includegraphics[width=\textwidth,valign=t]{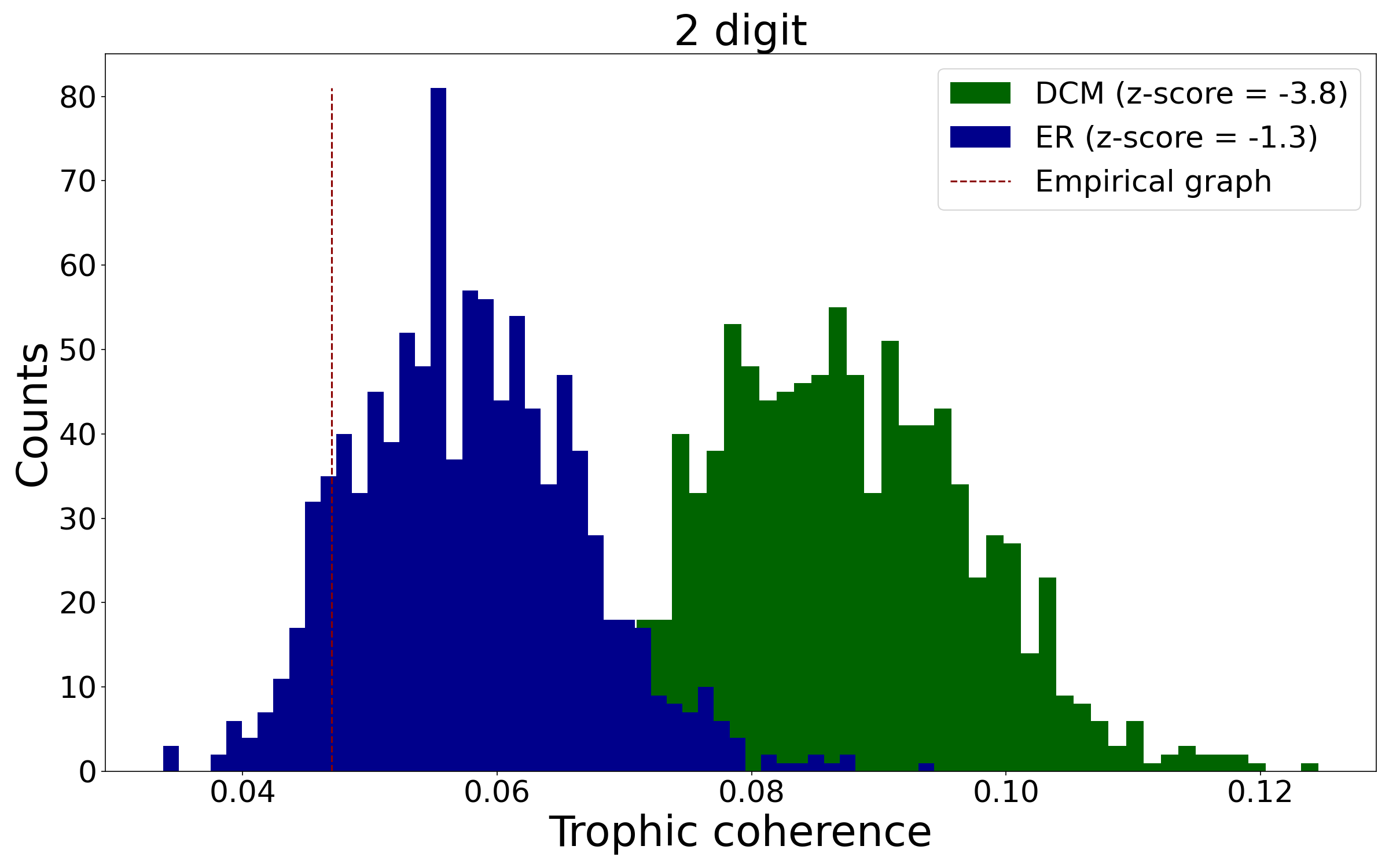}
    \end{subfigure}
    \caption{Empirical value (vertical red line) and null distributions of the trophic coherence of the input-output networks built at 6 digit (\textbf{a}), 4 digit (\textbf{b}) and 2 digit (\textbf{c}) aggregation level. The trophic coherence is significant with respect to both the \emph{Erdòs-Renyi} model and the \emph{Directed Configuration Model} only for products at the 6-digit level (\textbf{a}), signaling that the trophic organization of the product's network emerges only at a very detailed scale. The null distributions are obtained from 1000 explicit samples of the null models.}
    \label{fig:troph_coh}
\end{figure}

Although the trophic coherence of our reconstructed value chain is quite low ($TC = 0.15$), it is statistically significant - as shown in figure \ref{fig:troph_coh}a - with respect to the distributions induced by two different null models: the \emph{Erdòs-Renyi} (ER) random graph \cite{erdds1959random}, preserving the network's density, and the \emph{Directed Configuration Model} \cite{squartini2011analytical} (DCM), preserving the (in- and out-) degree sequences of nodes. This result suggests that the input-output network does possess a fundamental trophic organization, as qualitatively confirmed by the trophic level rankings in table \ref{tab:troph_lev}. This structure is however blurred by the circularity of production processes emerging at the firm level: while there is a clear distinction between basic goods (e.g. textile fibres or animals) and final goods (e.g. vehicles), the production of the former still needs the input of many complex products (e.g. machinery, packaging tools and such).

In order to inspect the effect of products' aggregation level on the trophic structure of the network, we repeated the analysis for products grouped at the 4- and 2-digit level, following the HS taxonomy (see supplementary section S1). Note that, differently from fig.\ref{fig:sect_net}, in this case the aggregation of products was performed on the raw import/export data, i.e. before the network building procedure (described in the Methods section). The results show that at the 4-digit level (fig.\ref{fig:troph_coh}b) the trophic coherence of the network is compatible with the distribution of the DCM, albeit still significant with respect to the ER model, while at the 2-digit level (fig.\ref{fig:troph_coh}c) it is compatible with the null models, being lower than both: in other words, at this level of aggregation, random networks with the same density or degree sequence as the empirical one have a more pronounced trophic structure. We can thus conclude that the relatively weak, although significant, trophic organization of the products' input-output network is present at a very granular scale, but it is lost upon aggregating products into coarser sectors. From this observation, it is clear how the picture that can be obtained with the data and methods presented in this paper is not only more detailed, but also qualitatively different from the one usually obtained from the analysis of coarse-grained international input-output tables.

\subsection*{Firms' trophic direction}

As the trophic hierarchy of products reflects the transformations induced by productive processes, it is natural to inspect whether firms export, on average, products with a higher trophic level than their imports, i.e. if they have a positive \emph{trophic jump} (see \cref{eq:import_tl,eq:export_tl,eq:tj}).

\begin{figure}[!ht]
    \centering
    \includegraphics[width=0.7\linewidth]{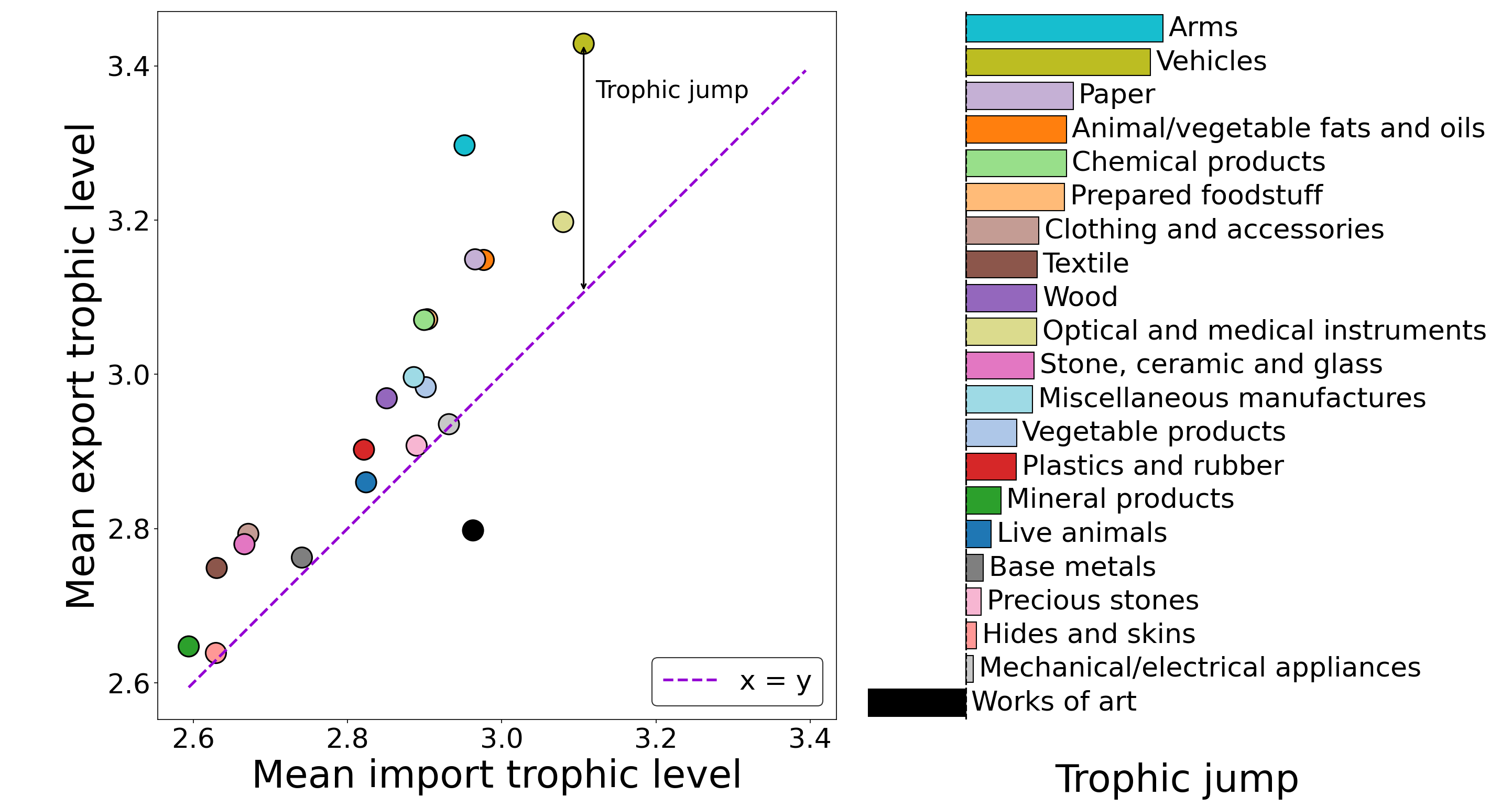}
    \caption{Average export vs import trophic level of firms grouped into industrial sectors. The distance from the identity line (\emph{trophic jump}) measures the transformation operated by industries on their inputs, while their position along the identity line indicates at which stage of the trophic hierarchy they operate. The barplot shows the magnitude of the \emph{trophic jump} for every sector.}
    \label{fig:firms_tl}
\end{figure}

Looking at the binarized import and export matrices for 2007 (\cref{eq:rca,eq:rca_filter}), the same year used to build the products' network, among the $19.294$ firms that do not import and export the same product (among others), $59.7\%$ display a positive trophic jump, while the opposite is true for the remaining $40.3\%$: trophic jump values follow a normal distribution centered around $\simeq0.1$ (see supplementary figure S2). In order to inspect whether the trophic jump is related to the industrial sector, following the HS classification (see supplementary section S1), we grouped firms at the \textit{sections} level (see also fig. \ref{fig:sect_net}), based on their main exported product in terms of monetary volume. All sectors display a percentage of firms with positive trophic jump above $50\%$, except for \emph{Works of art} and \emph{Live animals}: the percentage is lower for sectors closer to primary industries, as \emph{Live animals} and \emph{Hides and skins}, and is highest for technology-intensive industries, as \emph{Arms and ammunition} and \emph{Vehicles} (see supplementary figure S3).

For all sectors, however, firms export, on average, products with a higher trophic level than the ones they import, with the sole exception of \emph{Works of art}, as displayed in figure \ref{fig:firms_tl}. The highest trophic jumps are realized by \emph{Arms and ammunition} and \emph{Vehicles}, i.e. industries involving a substantial transformation of their inputs, while the lowest belong to industries involved in the production of raw materials, e.g. \emph{Precious stones} and \emph{Hides and skins}, with the exception of \emph{Mechanical and electrical appliances}. 
The position of industries along the identity line provides information on which stage of the trophic hierarchy they occupy. Industries in the bottom left corner group firms that, on average, import and export products with low trophic level, i.e. close to primary goods, as \emph{Mineral products} and \emph{Hides and skins}; industries on the upper right corner, on the other hand, operate closer to final consumption, as \emph{Vehicles} and \emph{Optical and medical instruments}.

\subsection*{Countries' imports and exports}

\begin{figure}[!ht]
    \centering
    \includegraphics[width=0.7\linewidth]{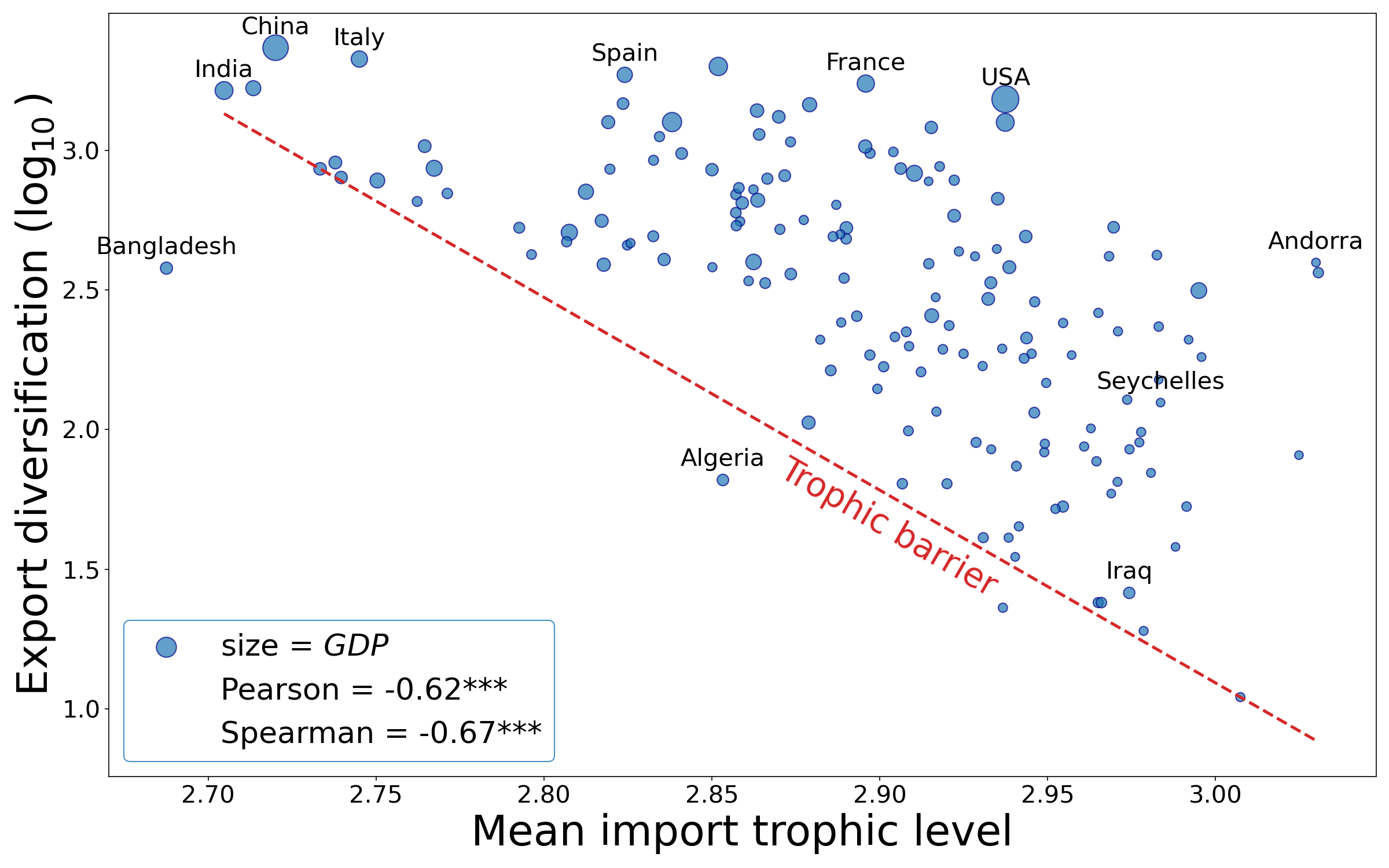}
    \caption{Export diversification of countries vs average trophic level of their imports. There is a strong and significant negative correlation between the two quantities, both in terms of values (Pearson) and rankings (Spearman). Asterisks denote the significance level of the correlation coefficients: $p\leq0.01$ (***). The plot exhibits a triangular structure, with the bottom left corner completely empty. This means that only developed countries can choose their import strategy, on the basis of the respective industrial policies and position in the global supply chain.}
    \label{fig:div_vs_imp_tl}
\end{figure}

\noindent We now focus our analysis on countries' imports and exports, to inspect whether countries with different industrial capabilities operate at different stages of the trophic hierarchy.

\begin{figure}[!ht]
    \centering
    \begin{minipage}{0.6\linewidth}
        \centering
        \includegraphics[width=\linewidth]{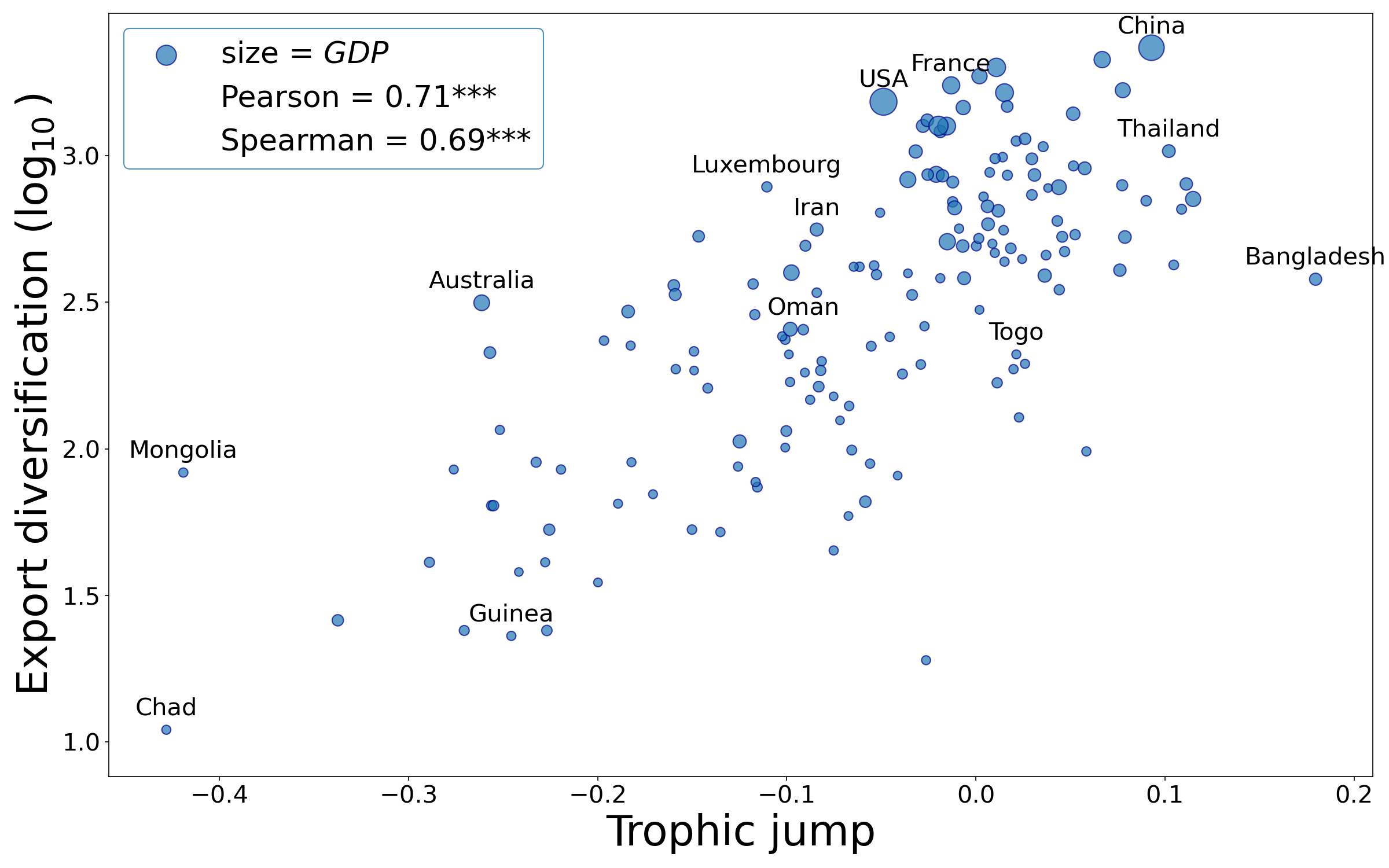}
    \end{minipage}%
    \hfill
    \begin{minipage}{0.38\linewidth}
        \centering
        \begin{tabular}{lcc}
            \toprule
            & NTJ & Ranking \\
            \midrule
            Bangladesh & 0.0303 & 1 \\
            Mexico & 0.0175 & 2  \\
            Syria & 0.0174 & 3  \\
            ... &  & \\
            Croatia & 0.0015 & 51\\
            Germany & 0.0014 & 52 \\
            Moldova & 0.0014 & 53 \\
            ... &  &  \\
            Guinea-Bissau & -0.0664 &  161\\
            Greenland & -0.0699 &  162\\
            Angola & -0.0714 &  163 \\
            \bottomrule
        \end{tabular}
    \end{minipage}
    \caption{Left: export diversification of countries vs average trophic jump. There is a strong and significant positive correlation between the two quantities, both in terms of values (Pearson) and rankings (Spearman). Asterisks denote the significance level of the correlation coefficients: $p\leq0.01$ (***). Right: table reporting the value of the normalized trophic jump (NTJ) - defined as trophic jump divided by the logarithm of export diversification - and the corresponding ranking for top, middle, and bottom countries.}
    \label{fig:div_vs_delta_tl}
\end{figure}

Employing 2023 import/export data provided by UN-COMTRADE we build the binarized import and export matrices $M^{I}$ and $M^{E}$, whose elements $m_{cp}$ are equal to 1 if country c competitively exports/imports product p (see Methods section). This allows us to compute the average trophic level of each country's imports and exports (see \cref{eq:import_tl,eq:export_tl}).

The scatterplot of export diversification versus average import trophic level of countries (figure \ref{fig:div_vs_imp_tl}) displays a strong, negative correlation, pointing out that more diversified countries (i.e. more industrially proficient) import, on average, more upstream products, and vice versa. This result confirms the validity of the information embedded in trophic levels, as it is natural to expect that countries with greater ability to transform products through industrial processes tend to import basic goods to input to their production, while on the other hand countries with low industrial capacity need to import final products that they are not able to produce. A more detailed inspection of the plot reveals a triangular relationship between the two quantities, highlighted by the dotted line: while the trend is overall decreasing, still there are several highly diversified countries (i.e. France, and most manifestly USA) with high mean import trophic level, while there are no countries with low diversification importing (on average) upstream products, as shown by the absence of dots in the bottom left corner of the figure. This signals that for industrially proficient countries the trophic position of their imports, while overall tends to be upstream, still can be determined by other factors, such as their industrial specialization patterns (heavily manufacturing countries such as Italy, China and Bangladesh have lower average import trophic level than more service-oriented ones as USA and France), as well as their reliance on the import of natural resources. 
Industrially developing countries, on the other hand, are forced to import downstream products, as they do not have the capabilities to transform upstream goods, a constraint we call \emph{trophic barrier}. Oil exporters lie at the bottom part of the trophic barrier, too. Remarkably, the same triangular shape is present when looking at imports from a fixed country (see supplementary section S4), confirming the robustness and generality of our findings. 

We next investigate the trophic jump of countries (see \cref{eq:tj}), in analogy with the sector level analysis of fig.\ref{fig:firms_tl}. As shown in fig.\ref{fig:div_vs_delta_tl},
the diversification of countries displays a strong positive correlation with their trophic jump, meaning that more diversified countries tend to export products that are more downstream than the ones they import. This finding is perfectly in line with the interpretation of the previous plot, as the trophic jump is connected the value that the industrial infrastructure of a country adds in the transformation of the inputs (imports) into outputs (exports): the greater the industrial capabilities of the country, the bigger the portion of trophic chain that its industrial processes are able to cover. A more detailed inspection of the plot shows that the trophic jump is particularly high for countries whose economy is centered on the manufacturing of physical goods (e.g. China, Thailand, and Bangladesh) while it is relatively lower, given their diversification, for service-oriented countries (as USA and Luxembourg). We recall that these analyses employ the trophic levels computed on the products' network built from 2007 data (see Methods): however the same results hold when using trophic levels averaged over the 5 years span 2007-2011, as well as countries' import and export data for years other than 2023 (see Supplementary Section S5).

\subsection*{Trophic levels predict economic growth}

Taken together, these results indicate that the trophic direction represents a powerful description of countries' specialization patterns. In this section, we will verify whether this information is also a driver of countries' industrial growth. In particular, we test the ability of both average import trophic level and trophic jump to predict the future growth of $GDP_{pc}$ of countries in a linear regression setting. Borrowing from the standard econometric literature on the subject\cite{wacker2024leveraging}, we build a baseline linear regression model to predict the logarithmic growth of $\text{GDP}_{pc}$ for a country after $\Delta$ years, $\log\left(\frac{\text{GDP}_{pc}(y+\Delta)}{\text{GDP}_{pc}(y)}\right)$, using as explanatory variables the $\text{GDP}_{pc}$ itself (measured in constant 2015 US$\$$), the population of the country (POP) and the volume of realized trade as percentage of its GDP \footnote{This data was collected from the open access World Bank repository \emph{World Development Indicators}, available at https://databank.worldbank.org/source/world-development-indicators} , in year $y$, for $y\in[2007,2023-\Delta]$

\begin{equation}
    \log\left(\frac{\text{GDP}_{pc}(y+\Delta)}{\text{GDP}_{pc}(y)}\right) = \alpha + \beta \log\left(\text{GDP}_{pc}(y)\right) + \gamma\log\left(\text{POP}(y)\right) + \delta\log\left(\text{TRD}(y)\right) + \zeta(y) +  \epsilon
\end{equation}

\noindent where $\zeta(y)$ represents the year fixed effect and $\epsilon$ is the residual error. Note that we introduce clustered errors for the observations relative to the same country in different years, and standardize all variables.

Given this baseline model, we include as additional covariates the average import trophic level and the trophic jump of each country, both normalized by the logarithm of export diversification (NITL and NTJ, respectively): this normalization allows us to account for size effects, capturing trophic characteristics per unit of diversification rather than in absolute terms. Due to the strong collinearity between export diversification and the normalized average import trophic level, which would affect the stability and interpretability of the results, we opt to test separately models including the former, and models including the latter. 

The results, reported in table \ref{tab:growth_reg}, show that the model including both trophic level based variables (TL3) achieves the highest predictive power, as measured by the adjusted $R^{2}$, with both variables being statistically significant at the $p\leq0.01$ threshold. Both coefficients are negative, indicating a negative association with future economic growth. 

For the normalized average import trophic level (NITL), the interpretation is straightforward: countries importing more upstream products per unit export tend to experience higher future growth rates, consistent with the intuition that more upstream imports reflect greater industrial capabilities, required to process inputs and generate economic return. The negative coefficient for the normalized trophic jump (NTJ), on the other hand, is a bit counterintuitive, as one would expect countries with bigger trophic jumps per unit export to grow faster. However, this finding can be interpreted as the consequence of three main aspects. First, the negative sign emerges when the normalized trophic jump is included alongside either the normalized average import trophic level or export diversification - which are themselves highly correlated - suggesting that, conditional on a country’s industrial capabilities, a lower trophic jump per unit export may indicate unexploited potential for future growth. Second, the absence of services in our dataset likely distorts the representation of value chains by omitting a substantial share of value creation, particularly in high-income economies. Third, a high trophic jump per unit diversification may reflect specialization in assembly-based strategies, positioning the country as a platform for foreign value chains - as suggested by the fact that Bangladesh and Mexico display the highest NTJ values (see fig.\ref{fig:div_vs_delta_tl}, right); this interpretation aligns with the evidence that highly developed countries tend to export broader sets of goods spanning all complexity levels \cite{tacchella2012new}, which in turn reduces their average trophic jump (e.g. Croatia and Germany show values of NTJ close to 0, as shown in fig.\ref{fig:div_vs_delta_tl}, right). The same conclusions also hold when performing the regression to predict the growth of $\text{GDP}_{pc}$ after $\Delta=5$ years, and upon substituting trophic levels from 2007 data with averages over the 2007–2011 period (see Supplementary Section S6). 

As a robustness check, we address the collinearity between NITL, NTJ, and export diversification by orthogonalizing the latter two with respect to NITL. Specifically, we regress NTJ and export diversification separately on NITL, and use the residuals — representing the components orthogonal to NITL — as covariates in the growth regression\cite{angrist2009mostly} (see supplementary section S7 for details). The results, reported in supplementary table S4, confirm our main findings: both NITL and the residualized NTJ remain significant and negatively associated with growth, while export diversification adds no predictive power once NITL is accounted for (its residual is not significant).

\begin{table}[ht!]
\centering
\begin{tabular}{lllllll}
\hline
 & Baseline & TL1 & TL2 & TL3 & EXP & ETL2\\
\hline
$\text{GDP}_{pc}$ (log)      & -0,045*** & -0,071*** & -0.051***  & -0,075***          & -0,088*** & -0.094*** \\
Population (log)             & 0,050***  &  0,020    & 0.041**    &  0,018             & -0,001    & -0.005    \\
Trade \% of GDP (log)        & 0,057***  &  0,048*** & 0.051***   &  0,050***          & 0,041**   & 0.040**   \\
Constant                     & 0,155***  &  0,165*** & 0.161***   &  0,163***          & 0,163***  & 0.161***  \\
NITL              &           & -0,078*** &            & \textbf{-0,115***} &           &           \\
NTJ               &           &           & 0.035**    & \textbf{-0,045***} &           & -0.023*   \\
Exp div (log)                &           &           &            &                    &  0,096*** & 0.116***  \\
\hline
Adj. $R^{2}$                 & 0,075     & 0,198     & 0.105      & \textbf{0,218}     & 0,202     & 0.208     \\
N° observations              & 1.059     & 1.059     & 1.059      & 1.059              & 1.059     & 1.059     \\
\end{tabular}
\caption{Linear regression to predict $\log\left(\frac{\text{GDP}_{pc}(y+\Delta)}{\text{GDP}_{pc}(y)}\right)$, using a set of variables in year $y$, with $\Delta=10$ years. The model involving both trophic level-based covariates (TL3) displays the highest value of adjusted $R^{2}$, i.e., it is the most predictive one. All variables are standardized. Asterisks denote different statistical significance levels of the variables: $p\leq0.1$ (*), $p\leq0.05$ (**), $p\leq0.01$ (***).}
\label{tab:growth_reg}
\end{table}


\section*{Discussion}
In this work, we introduce a novel, data-driven approach to map and analyze the hierarchical structure of product-level value chains, leveraging detailed firm-level trade data and tools from ecological network theory. We construct a statistically validated input-output network of products by comparing the import and export of more than 170.000 Italian firms. Then, we define and compute the \textit{trophic level} of goods to quantify their production stage in the value chain. This methodology extends and refines traditional concepts such as upstreamness and downstreamness, revealing a clear, though nuanced, directional structure in the flow of goods — from basic inputs to complex, final products.

Our analysis shows that trophic levels provide a meaningful ordering of products that aligns with intuitive notions of value chain positioning. Importantly, we find that the \textit{trophic coherence} — the extent to which this ordering holds at scale — is significant only at the most granular level of product classification. This underscores the limitations of using aggregated industry-level data in value chain studies and highlights the key importance of high-resolution, firm-level datasets.

By analyzing firms’ trade patterns, we demonstrate that most firms operate with a positive trophic jump, exporting products that are more downstream than those they import, with variations across sectors reflecting different degrees of value addition. Extending this analysis to the country level, we find that more diversified economies tend to import upstream goods and export downstream ones, suggesting that trophic metrics capture core features of industrial maturity and capability. This picture is however nuanced, and the relation that we observe between diversification and the average trophic level of imports suggests the existence of a \emph{trophic barrier} that prevents economies with small export diversification from being importers of upstream goods. The opposite, however, is not true, and we observe highly diversified exporters that are also importers of downstream goods. The main outlier in this sense is the United States, and this finding is remarkable in light of the recent extreme tariff policy that the country is putting forward. Our findings demonstrate quantitatively and systematically that the US economy is operating at the very opposite end of the spectrum with respect to strong manufacturing economies with similar diversification levels, such as China, India, and Italy.

Finally, we show that trophic indicators not only provide a description of countries' industrial orientation, but they are also predictive. Both the average import trophic level and the trophic jump of a country are strong predictors of future economic growth. This result points to the central role of a country's position and progression along the value chain in shaping long-term development trajectories. At the same time, this provides a strong validation for our approach. 

In theory, our methodology can potentially be biased by two factors: being based entirely on Italian data, and considering only imports and exports of Italian firms, having no picture of the internal market. However, the fact that the trophic classification of products derived from this approach allows us to define metrics that predict growth at the global level is a strong indication that the patterns that we see are general enough and capture a significant part of the intrinsic microscopic input-output structure of the globalized value chains.

\section*{Methods}

\subsection*{Firms' import/export data}

The ISTAT \textit{Commercio con l’estero} dataset (www.coeweb.istat.it) provides information on yearly bilateral import and export flows between Italian firms and foreign countries, along with the details of the exchanged products for the period 1993-2017. Imported and exported goods are classified according to the \emph{Harmonized System}: each product is assigned a unique 6-digit code, with every two digits specifying a different aggregation level. To build the products' network, we focus on 2007, for which we have information on the import and export volumes of $5.127$ products for $170.021$ firms and $233$ origin/destination countries. We refer the reader to the supplementary section S1 for more details on the data structure and cleaning procedure.

\begin{figure}[!ht]
    \centering
    \includegraphics[width=0.7\linewidth]{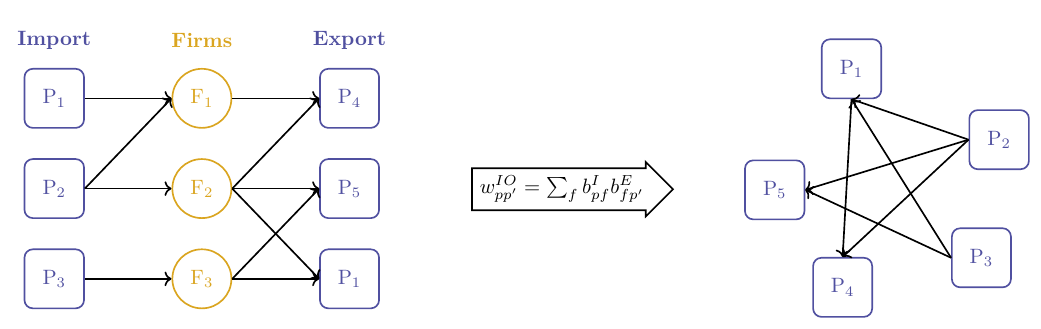}
    \caption{\textbf{Construction of the products' input-output network.} Given the tripartite network of imports and exports by Italian firms for a fixed yeary $y$, a directed monopartite network encoding input-output relationships between products is built by putting a link from product $p_{1}$ to product $p_{2}$ with weight equal to the number of firms that (competitively) import $p_{1}$ and simultaneously (competitively) export $p_{2}$.}
    \label{fig:network_construction}
\end{figure}

\subsection*{Constructing the products' network}

Upon aggregating, for every firm and product, the import and export volume from and to all countries of origin or destination, and performing a data cleaning procedure (described in supplementary section S1), we are left with $59.431$ firms and $5.008$ products. This import/export data can then be arranged into a tripartite network of products-firms-products, described by two bipartite adjacency matrices: $W^{I}$, whose entries $w^{I}_{pf}$ represent the (monetary) amount of product $p$ imported by firm $f$, and $W^{E}$, where $w^{E}_{fp}$ is the amount of product $p$ exported by firm $f$ (see fig.\ref{fig:network_construction}). Given this tripartite structure, we implement a two-step procedure to build a directed monopartite input-output network of products $W^{IO}$, with links $w^{IO}_{pp'}$ measuring how much product $p$ is needed as input in order to produce $p'$.

First, in order to discount the size of both products and firms, we replace the import (export) volume of each product $p$ by firm $f$ with the corresponding \emph{Revealed Comparative Advantage}\cite{balassa1965trade}

\begin{equation}
RCA^{I}_{pf} = \frac{w^{I}_{pf}/\sum_{p}w^{I}_{pf}}{\sum_{f}w^{I}_{pf}/\sum_{p,f}w^{I}_{pf}}
\label{eq:rca}
\end{equation}

\noindent
measuring the ratio between the weight product $p$ has in the total import of firm $f$ and the weight it has in the total import of all firms (and analogously on the export side). We then binarize the links by imposing the threshold $RCA_{pf}^{I}\geq1$, that distinguishes whether the import of product $p$ by firm $f$ is bigger than it is, on average, for other firms, or, in other words, if $f$ is a \emph{competitive} importer of product $p$ \cite{tacchella2012new}. Therefore we obtain two binary networks $B^{I}$ and $B^{E}$, with elements

\begin{equation}
    b^{I}_{pf} = \begin{cases}
        1 & \text{if}\,\,RCA^{I}_{pf} \geq 1\\
        0 & \text{otherwise}
    \end{cases}
\label{eq:rca_filter}
\end{equation}
\noindent and the same for $B^{E}$.

Then, we define the input-output networks of products $W^{IO}$ as

\begin{equation}
w^{IO}_{pp'} = \sum_{f}b^{I}_{pf}b^{E}_{fp'}
\label{eq:w_pp'}
\end{equation}

\noindent
where the link from product $p$ to $p'$ is given by the number of firms that import the former and simultaneously export the latter (see fig. \ref{fig:network_construction}), which acts as a proxy for the intensity of the input-output relationship between them. Self-loops, not providing any information on productive relationships between products, are eliminated from the graph.

\subsection*{Filtering the network}

This network construction procedure is prone to the presence of spurious links, due to two main sources of noise: ubiquitous imported (exported) products, which will be linked to many other products, without an actual productive relationship being present; and the presence of firms exporting multiple products, as all their imports will be linked to all their exports, despite possibly belonging to distinct production lines. In order to filter out these effects, keeping only those links signaling an actual productive dependence between products, we performed the statistical validation of the network against a suitable null model that accounts for the degrees of both products and firms.

As a first step we solved the \emph{Bipartite Configuration Model} (\emph{BiCM})\cite{saracco2015randomizing} separately on $B^{I}$ and $B^{E}$. This is an Exponential Random Graph model based on the maximum-entropy principle that, given an empirical bipartite network $B$, generates an ensemble of bipartite networks which are maximally random, except for the degrees of nodes on both layers, which are preserved as ensemble averages (we refer the reader to \cite{saracco2015randomizing} for the explicit derivation of the model). 

Therefore, for both the import and the export network, we obtained the corresponding random ensembles of graphs described by the link probabilities $p_{pf}^{I,BiCM}$ and $p_{fp}^{E,BiCM}$, respectively, satisfying 

\begin{align}
    &\langle k_{p}^{I}\rangle_{BiCM} = \sum_{f}p^{I,BiCM}_{pf} = \sum_{f}b^{I}_{pf} = k_{p}^{I}\\
    &\langle k_{f}^{I}\rangle_{BiCM} = \sum_{p}p^{I,BiCM}_{pf} = \sum_{p}b^{I}_{pf} = k_{f}^{I}\\
    &\langle k_{f}^{E}\rangle_{BiCM} = \sum_{p}p^{E,BiCM}_{fp} = \sum_{f}b^{E}_{pf} = k_{f}^{E}\\
    &\langle k_{p}^{E}\rangle_{BiCM} = \sum_{f}p^{E,BiCM}_{fp} = \sum_{p}b^{E}_{pf} = k_{p}^{E}
\end{align}

\noindent for every firm $f$ and product $p$. 

Since links in the \emph{BiCM} induced ensembles are independent random variables, the probability of a firm $f$ simultaneously importing product $p$ and exporting product $p'$ can be seen as the outcome of a Bernoulli trial with probability

\begin{equation}
    f(b^{I}_{pf}b^{E}_{fp'}) = p^{I,BiCM}_{pf}p^{E,BiCM}_{fp'}
\end{equation}

\noindent Therefore the link $w^{IO}_{pp'}$, given by the sum over all firms of these Bernoulli trials (\cref{eq:w_pp'}), is described in the random ensemble by a Poisson-Binomial distribution\cite{saracco2017inferring} $f_{PB}(w^{IO}_{pp'})$ with expected value

\begin{equation}
    \langle w^{IO}_{pp'}\rangle = \sum_{f}p^{I,BiCM}_{pf}p^{E,BiCM}_{fp'}
\end{equation}

\noindent We can then evaluate the statistical significance of the empirical link weights $w^{IO}_{pp'}$ by computing their p-values in the corresponding distribution as

\begin{equation}
    p-value(w^{IO}_{pp'}) = \sum_{x\geq w^{IO}_{pp'}}f_{PB}(x)
\end{equation}

Given the massive computational cost of estimating $N^{2}$ Poisson-Binomial distributions (where $N\simeq5000$ is the number of products), we performed an explicit sampling of random networks from the null ensemble (1000 samples), and estimated the p-values of the links empirically. 

Finally, to validate the p-values, we implemented the Benjamini-Hochberg procedure, correcting for multiple hypotheses testing\cite{thissen2002quick}. First we sorted the $M=N^{2}$ p-values in ascending order,

\begin{equation*}
    p_{1} \leq p_{2} \leq ... \leq p_{M}
\end{equation*}

\noindent and selected the largest integer $k$ such that

\begin{equation*}
    p_{k} \leq \frac{k\alpha}{M}
\end{equation*}

\noindent where $\alpha$ is the single test validation threshold, set to $\alpha=0.01$. Then we kept only those links $w_{pp'}^{IO}$ with p-value $p(w_{pp'}^{IO})\leq p_{k}$ and obtained the binary network $B^{IO}$, containing $N=5008$ products. As a last step, we focused on the \emph{giant weakly connected component} (GWCC) of the graph, covering $4966$ products. In table \ref{tab:net_prop} we report the main network statistics before and after the filtering. 

\begin{table}[ht!]
\centering
\begin{tabular}{lccc}
\hline
 & GWCC & Link density & $\langle k\rangle$ \\
\hline
Original network ($W^{IO}$) & 5008  & $1.7\cdot10^{-1}$ & 845 \\
Filtered network ($B^{IO}$) & 4966  & $1.8\cdot10^{-2}$ & 91 \\
\hline
\end{tabular}
\caption{Basic network statistics for products' network before and after the filtering: size of the \emph{giant weakly connected component} (GWCC), link density, and average degree.}
\label{tab:net_prop}
\end{table}

\subsection*{Trophic levels}

We compute trophic levels according to the definition introduced in \cite{mackay2020directed}. Given a directed network $B$ (either binary or weighted), we can define the vectors

\begin{align}
    &\vec{v} = \vec{s}_{in} - \vec{s}_{out}\\
    &\vec{u} = \vec{s}_{in} + \vec{s}_{out}
\end{align}

\noindent where ${s}_{in,j}=\sum_{i}b_{ij}$ and ${s}_{out,i}=\sum_{j}b_{ij}$, and the symmetrized network Laplacian

\begin{equation}
    \Lambda = diag(\vec{u}) - B - B^{T} .
\label{eq:lapl}
\end{equation}

\noindent The nodes' trophic levels $\vec{h}$ are then computed by solving the equation

\begin{equation}
    \Lambda\vec{h} = \vec{v}
\label{eq:troph_lev}
\end{equation}

\noindent which has a unique solution up to an additive constant in every connected component of the graph (see \cite{mackay2020directed} for details). Given the vector of trophic levels, we can look at how trophic the overall structure of the network is by defining the \emph{trophic coherence}

\begin{equation}
    TC(B) = 1 - \frac{\sum_{i,j}b_{ij}(h_{i}-h_{j}-1)^{2}}{\sum_{i,j}b_{ij}}
\label{eq:troph_coh}
\end{equation}

\noindent i.e., one minus the average square difference between the trophic distance of connected nodes and 1. In a perfectly trophic network (e.g. a linear graph or a tree), for any two connected nodes $i$ and $j$ we have $\lvert h_{i}-h_{j}\rvert=1$, and consequently $TC=1$, while in a network with no trophic hierarchy at all (e.g., a fully connected network), $\lvert h_{i}-h_{j}\rvert=0$, i.e. $TC=0$.

Given firms' import and export matrices $B^{I}$ and $B^{E}$, and the vector of trophic levels $\vec{h}$, we can define, for a firm $f$, its average import and export trophic levels

\begin{align}
    ITL_{f} = \frac{\sum_{p}b^{I}_{pf}h_{p}}{\sum_{p}b^{I}_{pf}}\\
\label{eq:import_tl}
    ETL_{f} = \frac{\sum_{p}b^{E}_{fp}h_{p}}{\sum_{p}b^{E}_{fp}}\\
\label{eq:export_tl}
\end{align}

\noindent and its \emph{trophic jump}

\begin{equation}
    TJ_{f} = ITL_{f} - ETL_{f} 
\label{eq:tj}
\end{equation}

\subsection*{Countries' import/export data}

Data about countries' export is gathered from UN-COMTRADE (available at https://comtrade.un.org upon subscription), providing bilateral export flows between countries, with products classified according to the \emph{Harmonized System} (HS) at the 6-digit level. Import and export declarations are subjected to a Bayesian reconciliation procedure\cite{Tacchella:2018aa}, leading to the construction of the annual import and export matrices $I(y)$ and $E(y)$ , whose elements $i_{cp}(y)$ ($e_{cp}(y)$) represent the monetary value of the import (export) of product $p$ by country $c$ in year $y$, for 169 countries and $\simeq5053$ products (depending on the HS edition, see supplementary section S1). Starting from these weighted matrices we build the binary import and export matrices $M^{I}(y)$ and $M^{E}(y)$ by computing the \emph{Revealed Comparative Advantage} and imposing the threshold $RCA=1$ (see \cref{eq:rca,eq:rca_filter}). By focusing on the imports and exports from or to a fixed country $\overline{c}$, and following the exact same pipeline, we build the fixed country import and export matrices $M^{I}_{\overline{c}}(y)$ and $M^{E}_{\overline{c}}(y)$, used in supplementary section S4.


\bibliography{references}

\begin{thebibliography}{10}
\urlstyle{rm}
\expandafter\ifx\csname url\endcsname\relax
  \def\url#1{\texttt{#1}}\fi
\expandafter\ifx\csname urlprefix\endcsname\relax\def\urlprefix{URL }\fi
\expandafter\ifx\csname doiprefix\endcsname\relax\def\doiprefix{DOI: }\fi
\providecommand{\bibinfo}[2]{#2}
\providecommand{\eprint}[2][]{\url{#2}}

\bibitem{antras2012measuring}
\bibinfo{author}{Antr{\`a}s, P.}, \bibinfo{author}{Chor, D.}, \bibinfo{author}{Fally, T.} \& \bibinfo{author}{Hillberry, R.}
\newblock \bibinfo{journal}{\bibinfo{title}{Measuring the upstreamness of production and trade flows}}.
\newblock {\emph{\JournalTitle{American Economic Review}}} \textbf{\bibinfo{volume}{102}}, \bibinfo{pages}{412--416}, \doiprefix\url{https://doi.org/10.1257/aer.102.3.412} (\bibinfo{year}{2012}).

\bibitem{fally2012production}
\bibinfo{author}{Fally, T.}
\newblock \bibinfo{journal}{\bibinfo{title}{Production staging: measurement and facts}}.
\newblock {\emph{\JournalTitle{Boulder, Colorado, University of Colorado Boulder, May}}} \bibinfo{pages}{155--168} (\bibinfo{year}{2012}).

\bibitem{miller2017output}
\bibinfo{author}{Miller, R.~E.} \& \bibinfo{author}{Temurshoev, U.}
\newblock \bibinfo{journal}{\bibinfo{title}{Output upstreamness and input downstreamness of industries/countries in world production}}.
\newblock {\emph{\JournalTitle{International regional science review}}} \textbf{\bibinfo{volume}{40}}, \bibinfo{pages}{443--475}, \doiprefix\url{https://doi.org/10.1177/0160017615608095} (\bibinfo{year}{2017}).

\bibitem{mcnerney2022production}
\bibinfo{author}{McNerney, J.}, \bibinfo{author}{Savoie, C.}, \bibinfo{author}{Caravelli, F.}, \bibinfo{author}{Carvalho, V.~M.} \& \bibinfo{author}{Farmer, J.~D.}
\newblock \bibinfo{journal}{\bibinfo{title}{How production networks amplify economic growth}}.
\newblock {\emph{\JournalTitle{Proceedings of the National Academy of Sciences}}} \textbf{\bibinfo{volume}{119}}, \bibinfo{pages}{e2106031118}, \doiprefix\url{https://doi.org/10.1073/pnas.2106031118} (\bibinfo{year}{2022}).

\bibitem{bartolucci2023correlation}
\bibinfo{author}{Bartolucci, S.}, \bibinfo{author}{Caccioli, F.}, \bibinfo{author}{Caravelli, F.} \& \bibinfo{author}{Vivo, P.}
\newblock \bibinfo{journal}{\bibinfo{title}{Correlation between upstreamness and downstreamness in random global value chains}}.
\newblock {\emph{\JournalTitle{arXiv preprint arXiv:2303.06603}}} \doiprefix\url{https://doi.org/10.48550/arXiv.2303.06603} (\bibinfo{year}{2023}).

\bibitem{bartolucci2025upstreamness}
\bibinfo{author}{Bartolucci, S.}, \bibinfo{author}{Caccioli, F.}, \bibinfo{author}{Caravelli, F.} \& \bibinfo{author}{Vivo, P.}
\newblock \bibinfo{journal}{\bibinfo{title}{Upstreamness and downstreamness in input--output analysis from local and aggregate information}}.
\newblock {\emph{\JournalTitle{Scientific Reports}}} \textbf{\bibinfo{volume}{15}}, \bibinfo{pages}{2727}, \doiprefix\url{https://doi.org/10.1038/s41598-025-86380-6} (\bibinfo{year}{2025}).

\bibitem{leontief1986input}
\bibinfo{author}{Leontief, W.}
\newblock \emph{\bibinfo{title}{Input-output economics}} (\bibinfo{publisher}{Oxford University Press}, \bibinfo{year}{1986}).

\bibitem{acemoglu2012net}
\bibinfo{author}{Acemoglu, D.}, \bibinfo{author}{Carvalho, V.~M.}, \bibinfo{author}{Ozdaglar, A.} \& \bibinfo{author}{Tahbaz-Salehi, A.}
\newblock \bibinfo{journal}{\bibinfo{title}{The network origins of aggregate fluctuations}}.
\newblock {\emph{\JournalTitle{Econometrica}}} \textbf{\bibinfo{volume}{80}}, \bibinfo{pages}{1977--2016}, \doiprefix\url{https://doi.org/10.3982/ECTA9623} (\bibinfo{year}{2012}).

\bibitem{inoue2019firm}
\bibinfo{author}{Inoue, H.} \& \bibinfo{author}{Todo, Y.}
\newblock \bibinfo{journal}{\bibinfo{title}{Firm-level propagation of shocks through supply-chain networks}}.
\newblock {\emph{\JournalTitle{Nature Sustainability}}} \textbf{\bibinfo{volume}{2}}, \bibinfo{pages}{841--847}, \doiprefix\url{https://doi.org/10.1038/s41893-019-0351-x} (\bibinfo{year}{2019}).

\bibitem{diem2022quantifying}
\bibinfo{author}{Diem, C.}, \bibinfo{author}{Borsos, A.}, \bibinfo{author}{Reisch, T.}, \bibinfo{author}{Kert{\'e}sz, J.} \& \bibinfo{author}{Thurner, S.}
\newblock \bibinfo{journal}{\bibinfo{title}{Quantifying firm-level economic systemic risk from nation-wide supply networks}}.
\newblock {\emph{\JournalTitle{Scientific reports}}} \textbf{\bibinfo{volume}{12}}, \bibinfo{pages}{1--13}, \doiprefix\url{https://doi.org/10.1038/s41598-022-11522-z} (\bibinfo{year}{2022}).

\bibitem{fessina2024inferring}
\bibinfo{author}{Fessina, M.} \emph{et~al.}
\newblock \bibinfo{journal}{\bibinfo{title}{Inferring firm-level supply chain networks with realistic systemic risk from industry sector-level data}}.
\newblock {\emph{\JournalTitle{arXiv preprint arXiv:2408.02467}}} \doiprefix\url{https://doi.org/10.48550/arXiv.2408.02467} (\bibinfo{year}{2024}).

\bibitem{inoue2023simulation}
\bibinfo{author}{Inoue, H.}, \bibinfo{author}{Okumura, Y.}, \bibinfo{author}{Torayashiki, T.} \& \bibinfo{author}{Todo, Y.}
\newblock \bibinfo{journal}{\bibinfo{title}{Simulation of supply chain disruptions considering establishments and power outages}}.
\newblock {\emph{\JournalTitle{PloS one}}} \textbf{\bibinfo{volume}{18}}, \bibinfo{pages}{e0288062}, \doiprefix\url{https://doi.org/10.1371/journal.pone.0288062} (\bibinfo{year}{2023}).

\bibitem{pichler2022forecasting}
\bibinfo{author}{Pichler, A.}, \bibinfo{author}{Pangallo, M.}, \bibinfo{author}{del Rio-Chanona, R.~M.}, \bibinfo{author}{Lafond, F.} \& \bibinfo{author}{Farmer, J.~D.}
\newblock \bibinfo{journal}{\bibinfo{title}{Forecasting the propagation of pandemic shocks with a dynamic input-output model}}.
\newblock {\emph{\JournalTitle{Journal of Economic Dynamics and Control}}} \textbf{\bibinfo{volume}{144}}, \bibinfo{pages}{104527} (\bibinfo{year}{2022}).

\bibitem{ialongo2022reconstructing}
\bibinfo{author}{Ialongo, L.~N.} \emph{et~al.}
\newblock \bibinfo{journal}{\bibinfo{title}{Reconstructing firm-level interactions in the dutch input--output network from production constraints}}.
\newblock {\emph{\JournalTitle{Scientific Reports}}} \textbf{\bibinfo{volume}{12}}, \bibinfo{pages}{1--12}, \doiprefix\url{https://doi.org/10.1038/s41598-022-13996-3} (\bibinfo{year}{2022}).

\bibitem{mungo2024reconstructing}
\bibinfo{author}{Mungo, L.}, \bibinfo{author}{Brintrup, A.}, \bibinfo{author}{Garlaschelli, D.} \& \bibinfo{author}{Lafond, F.}
\newblock \bibinfo{journal}{\bibinfo{title}{Reconstructing supply networks}}.
\newblock {\emph{\JournalTitle{Journal of Physics: Complexity}}} \textbf{\bibinfo{volume}{5}}, \bibinfo{pages}{012001}, \doiprefix\url{https://doi.org/10.1088/2632-072X/ad30bf} (\bibinfo{year}{2024}).

\bibitem{kosasih2021machine}
\bibinfo{author}{Kosasih, E.~E.} \& \bibinfo{author}{Brintrup, A.}
\newblock \bibinfo{journal}{\bibinfo{title}{A machine learning approach for predicting hidden links in supply chain with graph neural networks}}.
\newblock {\emph{\JournalTitle{International Journal of Production Research}}} \bibinfo{pages}{1--14}, \doiprefix\url{https://doi.org/10.1080/00207543.2021.1956697} (\bibinfo{year}{2021}).

\bibitem{wichmann2020extracting}
\bibinfo{author}{Wichmann, P.}, \bibinfo{author}{Brintrup, A.}, \bibinfo{author}{Baker, S.}, \bibinfo{author}{Woodall, P.} \& \bibinfo{author}{McFarlane, D.}
\newblock \bibinfo{journal}{\bibinfo{title}{Extracting supply chain maps from news articles using deep neural networks}}.
\newblock {\emph{\JournalTitle{International Journal of Production Research}}} \textbf{\bibinfo{volume}{58}}, \bibinfo{pages}{5320--5336}, \doiprefix\url{https://doi.org/10.1080/00207543.2020.1720925} (\bibinfo{year}{2020}).

\bibitem{chakraborty2024inequality}
\bibinfo{author}{Chakraborty, A.}, \bibinfo{author}{Reisch, T.}, \bibinfo{author}{Diem, C.}, \bibinfo{author}{Astudillo-Est{\'e}vez, P.} \& \bibinfo{author}{Thurner, S.}
\newblock \bibinfo{journal}{\bibinfo{title}{Inequality in economic shock exposures across the global firm-level supply network}}.
\newblock {\emph{\JournalTitle{Nature Communications}}} \textbf{\bibinfo{volume}{15}}, \bibinfo{pages}{3348}, \doiprefix\url{https://doi.org/10.1038/s41467-024-46126-w} (\bibinfo{year}{2024}).

\bibitem{bacilieri2022firm}
\bibinfo{author}{Bacilieri, A.}, \bibinfo{author}{Borsos, A.}, \bibinfo{author}{Astudillo-Estevez, P.} \& \bibinfo{author}{Lafond, F.}
\newblock \bibinfo{title}{Firm-level production networks: what do we (really) know?}
\newblock \bibinfo{type}{Tech. Rep.} \bibinfo{number}{2023-08}, \bibinfo{institution}{INET Oxford Working Paper} (\bibinfo{year}{2023}).

\bibitem{fessina2024pattern}
\bibinfo{author}{Fessina, M.}, \bibinfo{author}{Zaccaria, A.}, \bibinfo{author}{Cimini, G.} \& \bibinfo{author}{Squartini, T.}
\newblock \bibinfo{journal}{\bibinfo{title}{Pattern-detection in the global automotive industry: a manufacturer-supplier-product network analysis}}.
\newblock {\emph{\JournalTitle{Chaos, Solitons \& Fractals}}} \textbf{\bibinfo{volume}{181}}, \bibinfo{pages}{114630}, \doiprefix\url{https://doi.org/10.1016/j.chaos.2024.114630} (\bibinfo{year}{2024}).

\bibitem{brintrup2015nested}
\bibinfo{author}{Brintrup, A.}, \bibinfo{author}{Barros, J.} \& \bibinfo{author}{Tiwari, A.}
\newblock \bibinfo{journal}{\bibinfo{title}{The nested structure of emergent supply networks}}.
\newblock {\emph{\JournalTitle{IEEE Systems Journal}}} \textbf{\bibinfo{volume}{12}}, \bibinfo{pages}{1803--1812}, \doiprefix\url{https://doi.org/10.1109/JSYST.2015.2493345} (\bibinfo{year}{2018}).

\bibitem{diem2024estimating}
\bibinfo{author}{Diem, C.}, \bibinfo{author}{Borsos, A.}, \bibinfo{author}{Reisch, T.}, \bibinfo{author}{Kert{\'e}sz, J.} \& \bibinfo{author}{Thurner, S.}
\newblock \bibinfo{journal}{\bibinfo{title}{Estimating the loss of economic predictability from aggregating firm-level production networks}}.
\newblock {\emph{\JournalTitle{PNAS nexus}}} \textbf{\bibinfo{volume}{3}}, \bibinfo{pages}{pgae064}, \doiprefix\url{https://doi.org/10.1093/pnasnexus/pgae064} (\bibinfo{year}{2024}).

\bibitem{karbevska2025mapping}
\bibinfo{author}{Karbevska, L.} \& \bibinfo{author}{Hidalgo, C.~A.}
\newblock \bibinfo{journal}{\bibinfo{title}{Mapping global value chains at the product level}}.
\newblock {\emph{\JournalTitle{EPJ Data Science}}} \textbf{\bibinfo{volume}{14}}, \bibinfo{pages}{21}, \doiprefix\url{https://doi.org/10.1140/epjds/s13688-025-00521-5} (\bibinfo{year}{2025}).

\bibitem{albora2022machine}
\bibinfo{author}{Albora, G.} \& \bibinfo{author}{Zaccaria, A.}
\newblock \bibinfo{journal}{\bibinfo{title}{Machine learning to assess relatedness: the advantage of using firm-level data}}.
\newblock {\emph{\JournalTitle{Complexity}}} \textbf{\bibinfo{volume}{2022}} (\bibinfo{year}{2022}).

\bibitem{squartini2011analytical}
\bibinfo{author}{Squartini, T.} \& \bibinfo{author}{Garlaschelli, D.}
\newblock \bibinfo{journal}{\bibinfo{title}{Analytical maximum-likelihood method to detect patterns in real networks}}.
\newblock {\emph{\JournalTitle{New Journal of Physics}}} \textbf{\bibinfo{volume}{13}}, \bibinfo{pages}{083001}, \doiprefix\url{https//doi.org/10.1088/1367-2630/13/8/083001} (\bibinfo{year}{2011}).

\bibitem{cimini2019statistical}
\bibinfo{author}{Cimini, G.} \emph{et~al.}
\newblock \bibinfo{journal}{\bibinfo{title}{The statistical physics of real-world networks}}.
\newblock {\emph{\JournalTitle{Nature Reviews Physics}}} \textbf{\bibinfo{volume}{1}}, \bibinfo{pages}{58--71}, \doiprefix\url{https://doi.org/10.1038/s42254-018-0002-6} (\bibinfo{year}{2019}).

\bibitem{saracco2015randomizing}
\bibinfo{author}{Saracco, F.}, \bibinfo{author}{Di~Clemente, R.}, \bibinfo{author}{Gabrielli, A.} \& \bibinfo{author}{Squartini, T.}
\newblock \bibinfo{journal}{\bibinfo{title}{Randomizing bipartite networks: the case of the world trade web}}.
\newblock {\emph{\JournalTitle{Scientific Reports}}} \textbf{\bibinfo{volume}{5}}, \bibinfo{pages}{10595}, \doiprefix\url{https://doi.org/10.1038/srep10595} (\bibinfo{year}{2015}).

\bibitem{saracco2017inferring}
\bibinfo{author}{Saracco, F.} \emph{et~al.}
\newblock \bibinfo{journal}{\bibinfo{title}{Inferring monopartite projections of bipartite networks: an entropy-based approach}}.
\newblock {\emph{\JournalTitle{New Journal of Physics}}} \textbf{\bibinfo{volume}{19}}, \bibinfo{pages}{053022}, \doiprefix\url{https://doi.org/10.1088/1367-2630/aa6b38} (\bibinfo{year}{2017}).

\bibitem{cimini2022meta}
\bibinfo{author}{Cimini, G.}, \bibinfo{author}{Carra, A.}, \bibinfo{author}{Didomenicantonio, L.} \& \bibinfo{author}{Zaccaria, A.}
\newblock \bibinfo{journal}{\bibinfo{title}{Meta-validation of bipartite network projections}}.
\newblock {\emph{\JournalTitle{Communications Physics}}} \textbf{\bibinfo{volume}{5}}, \bibinfo{pages}{76} (\bibinfo{year}{2022}).

\bibitem{levine1980several}
\bibinfo{author}{Levine, S.}
\newblock \bibinfo{journal}{\bibinfo{title}{Several measures of trophic structure applicable to complex food webs}}.
\newblock {\emph{\JournalTitle{Journal of theoretical Biology}}} \textbf{\bibinfo{volume}{83}}, \bibinfo{pages}{195--207}, \doiprefix\url{https://doi.org/10.1016/0022-5193(80)90288-X} (\bibinfo{year}{1980}).

\bibitem{mackay2020directed}
\bibinfo{author}{MacKay, R.~S.}, \bibinfo{author}{Johnson, S.} \& \bibinfo{author}{Sansom, B.}
\newblock \bibinfo{journal}{\bibinfo{title}{How directed is a directed network?}}
\newblock {\emph{\JournalTitle{Royal Society open science}}} \textbf{\bibinfo{volume}{7}}, \bibinfo{pages}{201138}, \doiprefix\url{https://doi.org/10.1098/rsos.201138} (\bibinfo{year}{2020}).

\bibitem{tumminello2005tool}
\bibinfo{author}{Tumminello, M.}, \bibinfo{author}{Aste, T.}, \bibinfo{author}{Di~Matteo, T.} \& \bibinfo{author}{Mantegna, R.~N.}
\newblock \bibinfo{journal}{\bibinfo{title}{A tool for filtering information in complex systems}}.
\newblock {\emph{\JournalTitle{Proceedings of the National Academy of Sciences}}} \textbf{\bibinfo{volume}{102}}, \bibinfo{pages}{10421--10426}, \doiprefix\url{https://doi.org/10.1073/pnas.0500298102} (\bibinfo{year}{2005}).

\bibitem{erdds1959random}
\bibinfo{author}{Erdős, P.} \& \bibinfo{author}{Rényi, A.}
\newblock \bibinfo{journal}{\bibinfo{title}{On random graphs {I}}}.
\newblock {\emph{\JournalTitle{Publ. math. debrecen}}} \textbf{\bibinfo{volume}{6}}, \bibinfo{pages}{18}, \doiprefix\url{https://doi.org/10.5486/PMD.1959.6.3-4.12} (\bibinfo{year}{1959}).

\bibitem{wacker2024leveraging}
\bibinfo{author}{Wacker, K.~M.}, \bibinfo{author}{Beyer, R.~C.} \& \bibinfo{author}{Moller, L.~C.}
\newblock \bibinfo{title}{Leveraging growth regressions for country analysis}.
\newblock \bibinfo{type}{Tech. Rep.}, \bibinfo{institution}{The World Bank} (\bibinfo{year}{2024}).
\newblock \doiprefix\url{https://openknowledge.worldbank.org/entities/publication/c88bdfc7-d358-4fe4-8c79-43b0229d6747}.

\bibitem{tacchella2012new}
\bibinfo{author}{Tacchella, A.}, \bibinfo{author}{Cristelli, M.}, \bibinfo{author}{Caldarelli, G.}, \bibinfo{author}{Gabrielli, A.} \& \bibinfo{author}{Pietronero, L.}
\newblock \bibinfo{journal}{\bibinfo{title}{A new metrics for countries' fitness and products' complexity}}.
\newblock {\emph{\JournalTitle{Scientific reports}}} \textbf{\bibinfo{volume}{2}}, \bibinfo{pages}{723}, \doiprefix\url{https://doi.org/10.1038/srep00723} (\bibinfo{year}{2012}).

\bibitem{angrist2009mostly}
\bibinfo{author}{Angrist, J.~D.} \& \bibinfo{author}{Pischke, J.-S.}
\newblock \emph{\bibinfo{title}{Mostly harmless econometrics: An empiricist's companion}} (\bibinfo{publisher}{Princeton university press}, \bibinfo{year}{2009}).

\bibitem{balassa1965trade}
\bibinfo{author}{Balassa, B.}
\newblock \bibinfo{journal}{\bibinfo{title}{Trade liberalisation and “revealed” comparative advantage 1}}.
\newblock {\emph{\JournalTitle{The manchester school}}} \textbf{\bibinfo{volume}{33}}, \bibinfo{pages}{99--123}, \doiprefix\url{https://doi.org/10.1111/j.1467-9957.1965.tb00050.x} (\bibinfo{year}{1965}).

\bibitem{thissen2002quick}
\bibinfo{author}{Thissen, D.}, \bibinfo{author}{Steinberg, L.} \& \bibinfo{author}{Kuang, D.}
\newblock \bibinfo{journal}{\bibinfo{title}{Quick and easy implementation of the benjamini-hochberg procedure for controlling the false positive rate in multiple comparisons}}.
\newblock {\emph{\JournalTitle{Journal of Educational and Behavioral Statistics}}} \textbf{\bibinfo{volume}{27}}, \bibinfo{pages}{77--83}, \doiprefix\url{https://doi.org/10.3102/10769986027001077} (\bibinfo{year}{2002}).

\bibitem{Tacchella:2018aa}
\bibinfo{author}{Tacchella, A.}, \bibinfo{author}{Mazzilli, D.} \& \bibinfo{author}{Pietronero, L.}
\newblock \bibinfo{journal}{\bibinfo{title}{A dynamical systems approach to gross domestic product forecasting}}.
\newblock {\emph{\JournalTitle{Nature Physics}}} \textbf{\bibinfo{volume}{14}}, \bibinfo{pages}{861--865}, \doiprefix\url{https://doi.org/10.1038/s41567-018-0204-y} (\bibinfo{year}{2018}).

\end{thebibliography}

\section*{Acknowledgements}
The authors acknowledge the PRIN project No. 20223W2JKJ “WECARE”, CUP B53D23003880006, financed by the Italian Ministry of University and Research (MUR), Piano Nazionale Di Ripresa e Resilienza (PNRR), Missione 4 “Istruzione e Ricerca” - Componente C2 Investimento 1.1, funded by the European Union - NextGenerationEU.\\
The authors thank Giulio Cimini, Lorenzo Cresti and Giorgos Morakis for the fruitful discussions.

\section*{Author contributions statement}

Study conception and design: MF, AT, AZ. Data collection: AZ. Data analysis: MF. Discussion and interpretation of results: MF, AT, AZ. Draft manuscript preparation: MF, AT, AZ.

\section*{Competing interests}

The authors declare no competing interests.

\clearpage

\section*{\huge Supplementary Information}

\renewcommand{\thesection}{S\arabic{section}}
\renewcommand{\thetable}{S\arabic{table}}
\renewcommand{\thefigure}{S\arabic{figure}}

\bigskip

\bigskip

\section{Data cleaning}

The dataset analyzed in the study, provided by ISTAT, contains annual import and export transactions by Italian firms', along with the products and the origin and destination countries, in the period 1993-2017. Products are classified according to the \emph{Harmonized Description and Coding System} (HS), providing a unique 6 digits code for every good, with every two digits relative to a different aggregation level of the taxonomy, e.g.

\begin{itemize}
    \item \textbf{10}: Cereals
    \item \textbf{10.01}: Wheat and meslin
    \item \textbf{10.01.10}: Durum wheat
\end{itemize}

\noindent The number of products ranges from around 5.000 at the 6 digit level to around 100 at the 2 digit level, which can in turn be grouped into $21$ sections at the higher aggregation level.

The HS classification was updated in 1992, 1996, 2002, 2007, 2012 and 2017, and products' codes in the data are updated accordingly, resulting in discontinuities between different years of the dataset. In order to build the products' network, we focused on data for 2007 (as it is the first year of a five years span with coherent products codes in the data), covering around 170.021 firms and 5.127 products (at the 6 digit level), and performed a two-step cleaning procedure:

\begin{enumerate}
    \item we erased transactions relative to HS codes not belonging to the official classification: these extra codes were introduced by ISTAT to classify unidentified products;
    \item we retained only those firms appearing as both importers and exporters.
\end{enumerate}

\noindent Following these steps, we are left with 59.341 firms and 5.008 products, covering more than $90\%$ of the exchanged monetary volume.

\section{Trophic level distribution}

\begin{figure}[!ht]
    \centering
    \includegraphics[width=0.7\linewidth]{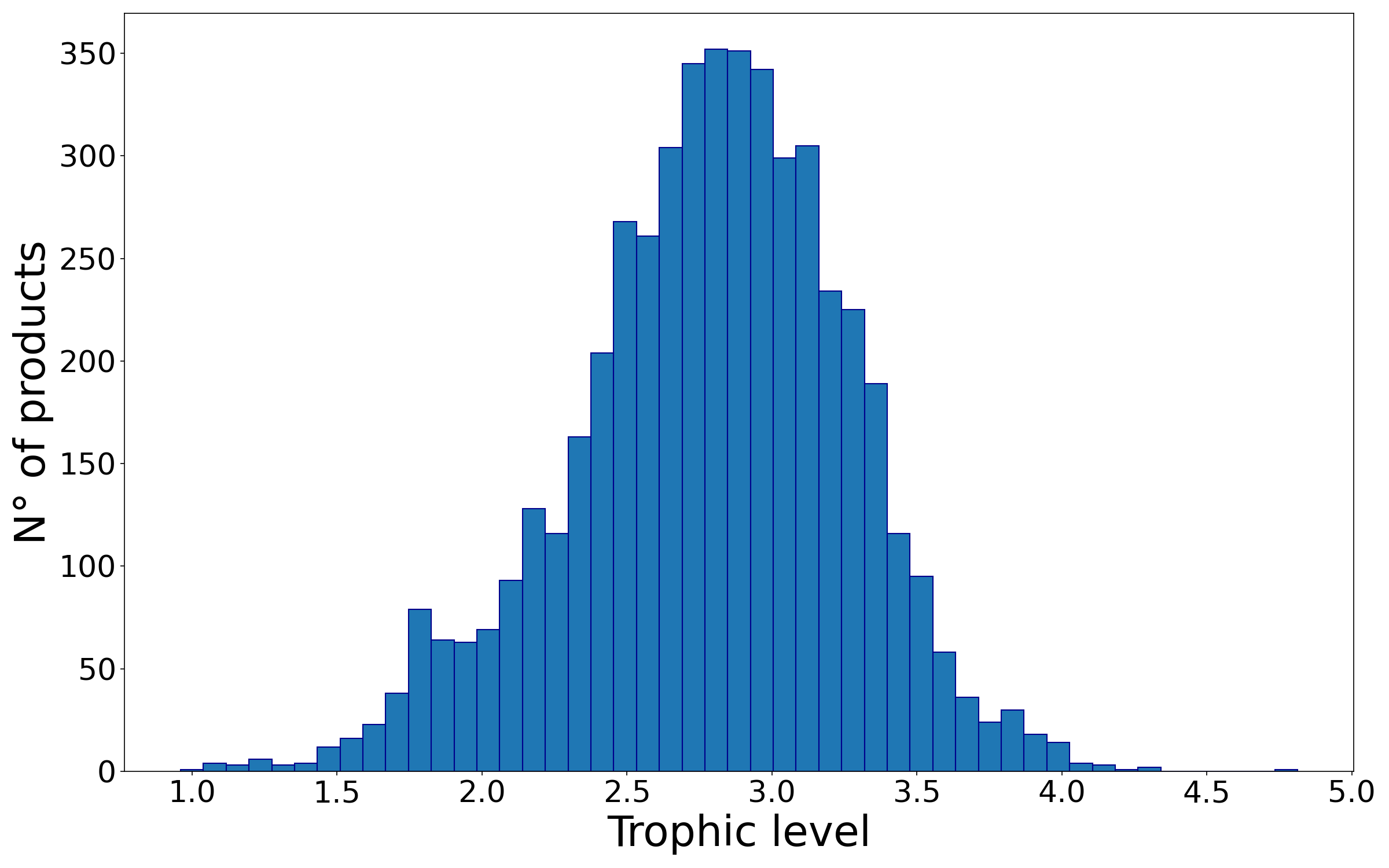}
    \caption{Empirical distribution of trophic levels in the network: the distribution resembles a right-skewed normal, with mean value $\overline{\text{TL}}=2.83$.}
    \label{fig:tl_distribution}
\end{figure}

In figure \ref{fig:tl_distribution} we report the empirical distribution of trophic levels in the network.

\section{Trophic jump distributions}

\begin{figure}[!hb]
    \centering
    \includegraphics[width=0.7\linewidth]{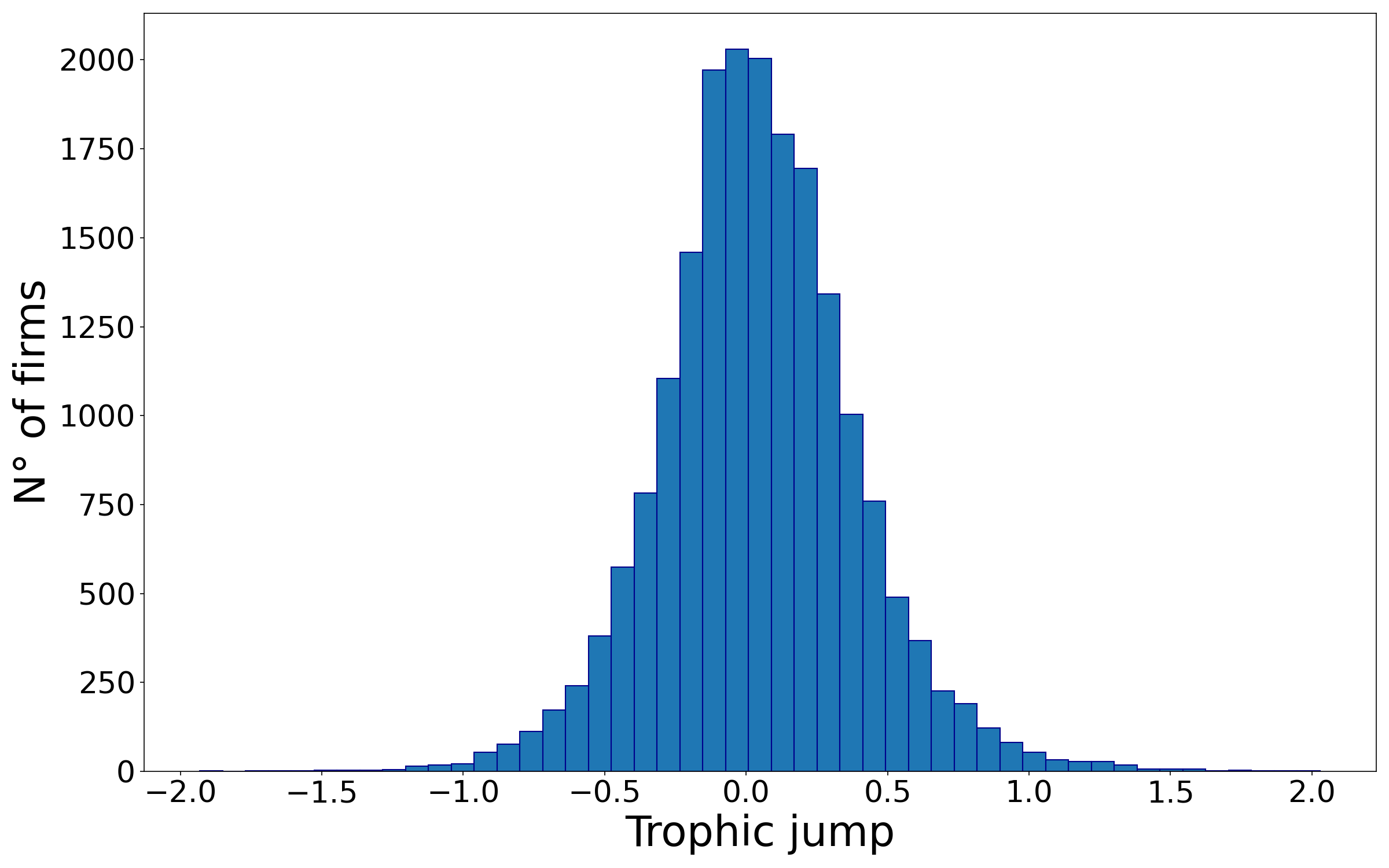}
    \caption{Empirical distribution of firms' trophic jumps: the distribution is normal, centered around the value $\simeq0.1$.}
    \label{fig:troph_dir_dis}
\end{figure}

In figure \ref{fig:troph_dir_dis} we report the empirical distribution of trophic jump for firms: as discussed in the main text, $50.9\%$ of firms have a positive trophic jump.

\begin{figure}[ht!]
    \centering
    \includegraphics[width=0.7\linewidth]{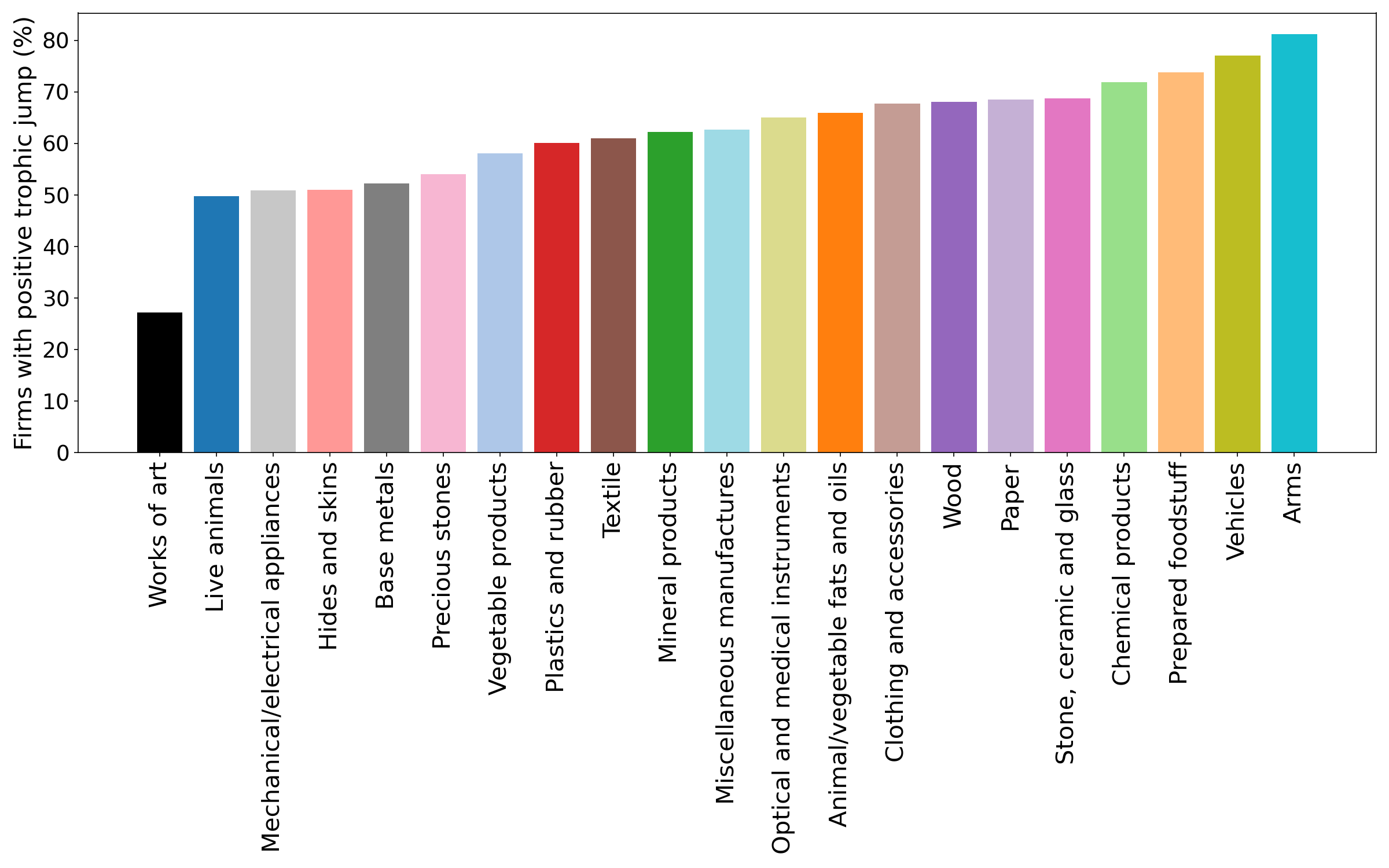}
    \caption{Percentage of firms with positive trophic jump in each sector. Sectors with the highest percentage are more technology intensive, as \emph{Vehicles} and \emph{Arms and ammunition}, while the lowest percentage belongs to industries involved in the production of primary goods (\emph{Live animals}, \emph{Base metals}), with the exception of \emph{Works of art}.}
    \label{fig:troph_dir_sect}
\end{figure}

Upon gathering firms into sector, all industries with the exception of \emph{Works of art} and \emph{Live animals} display a percentage of firms with positive trophic jump above $50\%$, as depicted in fig. \ref{fig:troph_dir_sect}.

\section{Average import trophic level from fixed countries}

\begin{figure}[b!]
    \centering
    \begin{subfigure}{0.02\textwidth}
    \textbf{a)}
    \end{subfigure}
    \begin{subfigure}[t]{0.47\textwidth}  \includegraphics[width=\textwidth,valign=t]{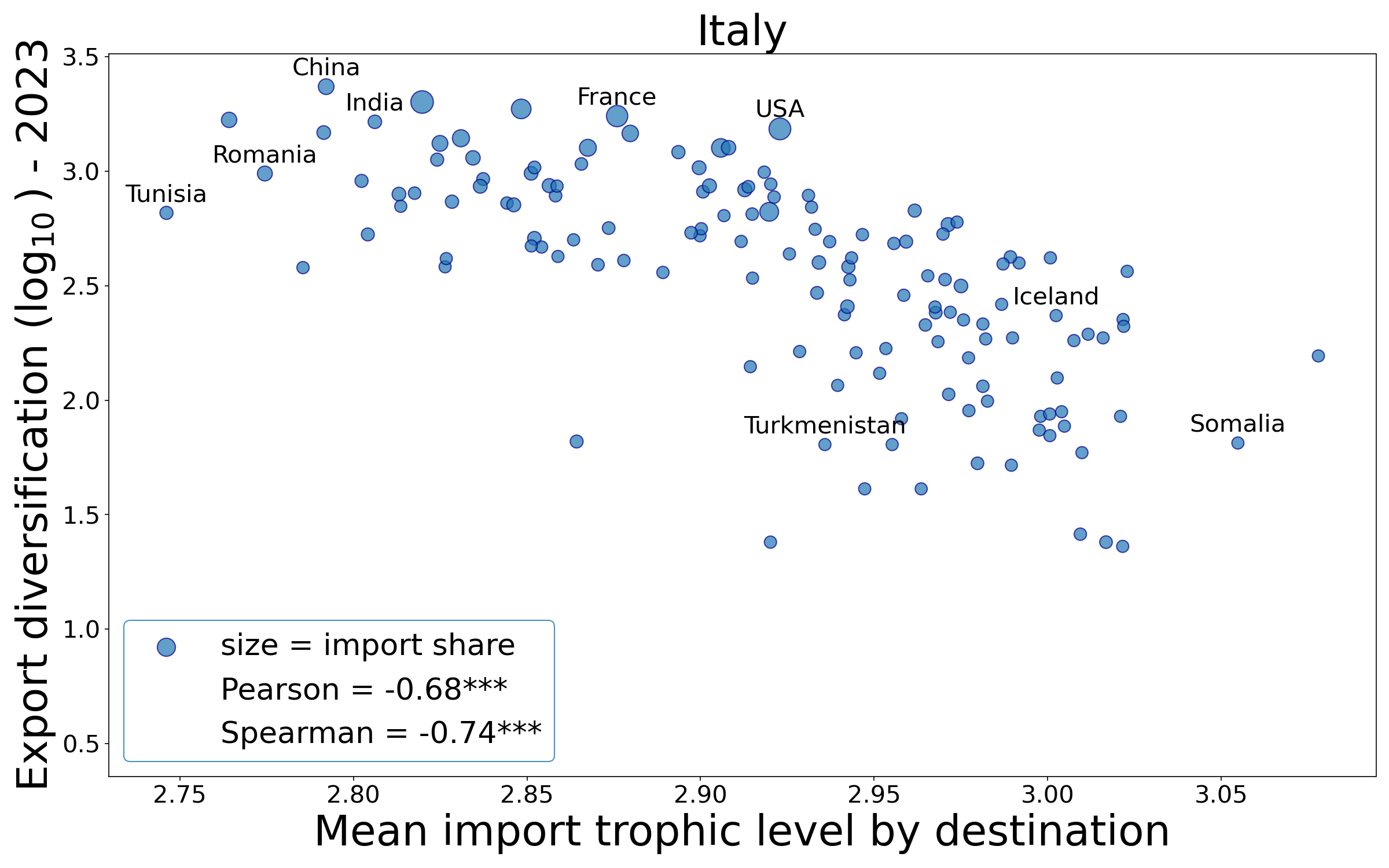}
    \end{subfigure}
    \begin{subfigure}{0.02\textwidth}
    \textbf{b)}
    \end{subfigure}
    \begin{subfigure}[t]{0.47\textwidth}  \includegraphics[width=\textwidth,valign=t]{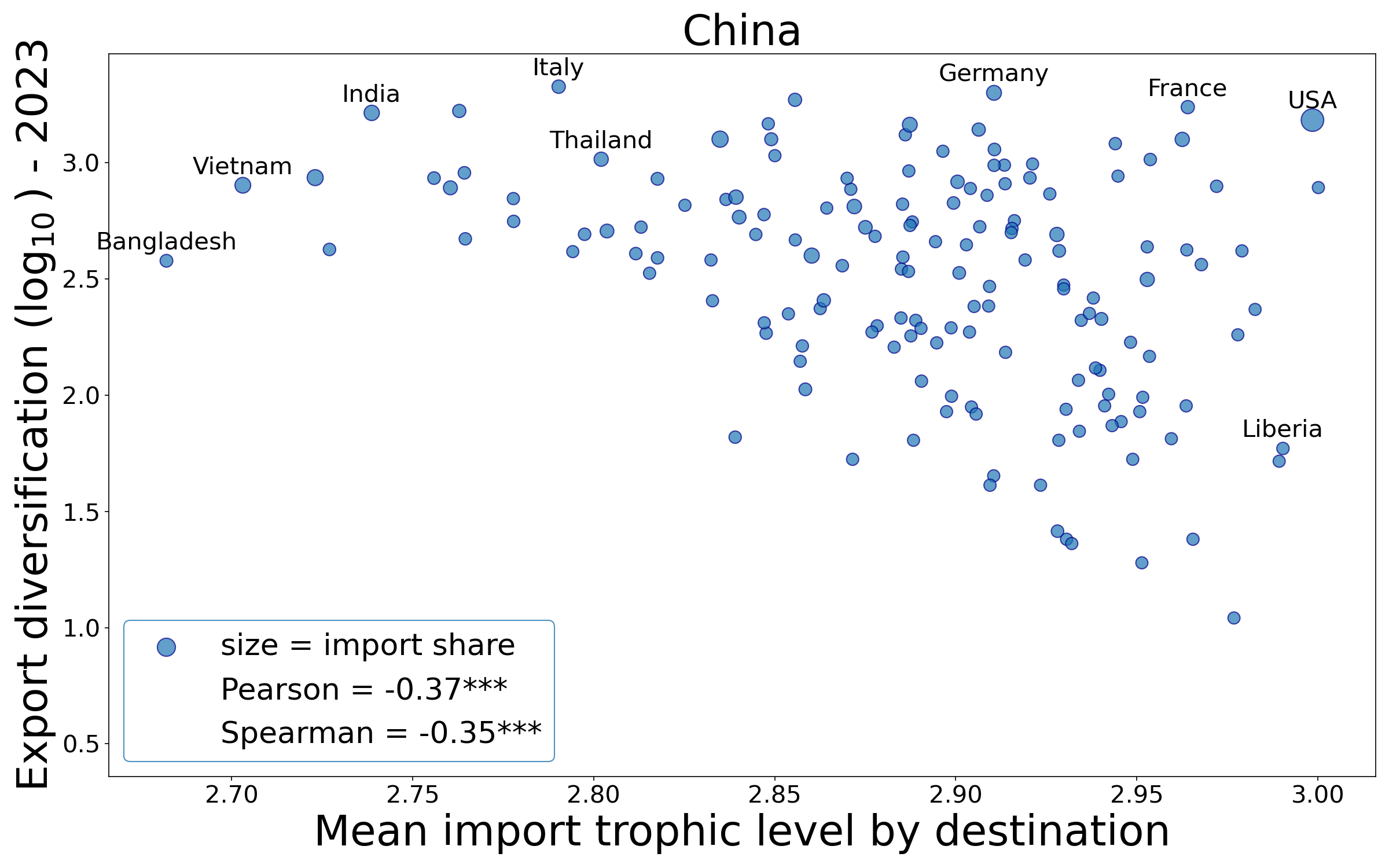}
    \end{subfigure}
    \begin{subfigure}{0.02\textwidth}
    \textbf{c)}
    \end{subfigure}
    \begin{subfigure}[t]{0.47\textwidth}  \includegraphics[width=\textwidth,valign=t]{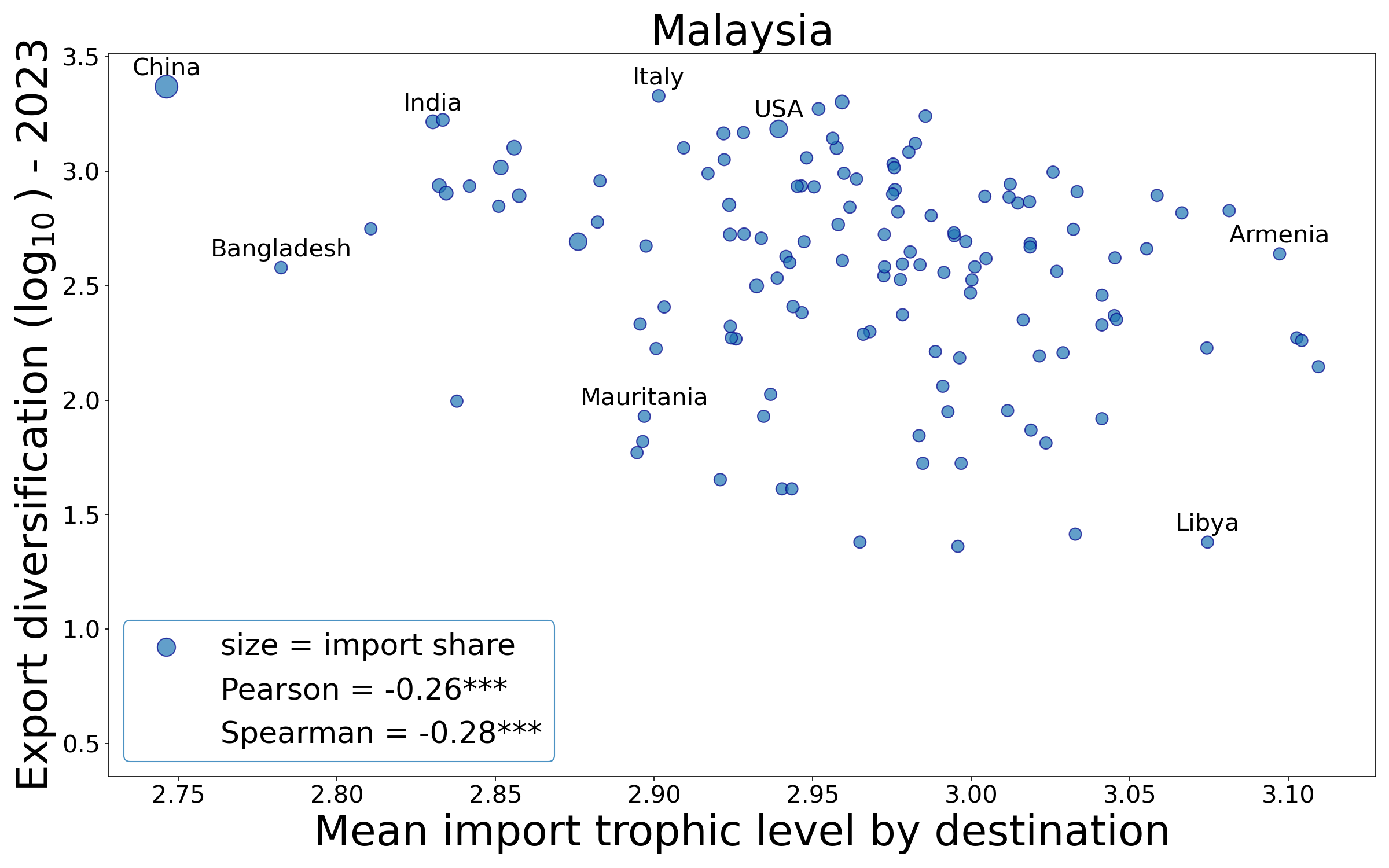}
    \end{subfigure}
    \begin{subfigure}{0.02\textwidth}
    \textbf{d)}
    \end{subfigure}
    \begin{subfigure}[t]{0.47\textwidth}  \includegraphics[width=\textwidth,valign=t]{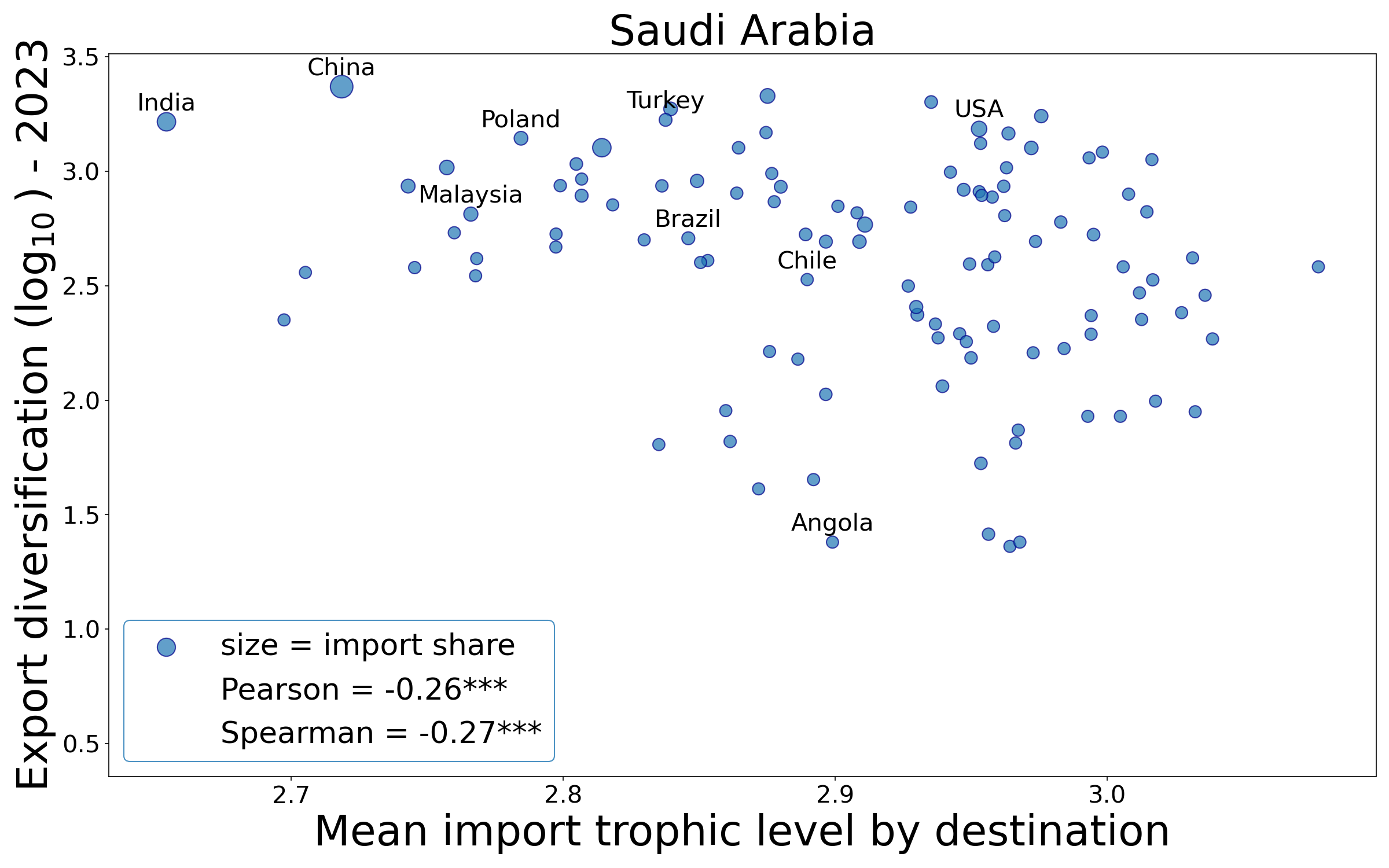}
    \end{subfigure}
    \caption{Export diversification versus average import trophic level for a fixed origin country: Italy (\textbf{a}), China (\textbf{b}), Malaysia (\textbf{c}) and Saudi Arabia (\textbf{d}). While only for Italy and China a strong, significant negative correlation between the two quantities is present, all plots display a triangular pattern, with no dots occupying the lower left corner. Asterisks denote the significance level of the correlation coefficients: $p\leq0.01$ (***)}
    \label{fig:div_vs_imp_tl_fc}
\end{figure}

We repeated the analysis on the average import trophic level of countries for a fixed origin country $\overline{c}$, i.e. by looking, for each country, at its total export diversification versus the average trophic level of products it imports from $\overline{c}$. In figure \ref{fig:div_vs_imp_tl_fc}a-d we show the plots for Italy, China, Malaysia and Saudi Arabia, respectively: while the negative correlation observed at the global level (see fig. 5 in the main text) is strong and significant only for Italy and for China, being lost for the other two countries, the triangular pattern is displayed by all plots, confirming that poorly diversified country do not import upstream products, neither globally nor from specific countries.

\section{Temporal robustness of countries' analysis}

In this section we test the robustness of our findings on countries' imports and exports with respect to a) changes in the year of import/export data for countries and b) changes in the year of firms' import/export data used to build the products' network on which trophic levels are computed.

\begin{figure}[t!]
    \centering
    \begin{subfigure}{0.02\textwidth}
    \textbf{a)}
    \end{subfigure}
    \begin{subfigure}[t]{0.47\textwidth}  \includegraphics[width=\textwidth,valign=t]{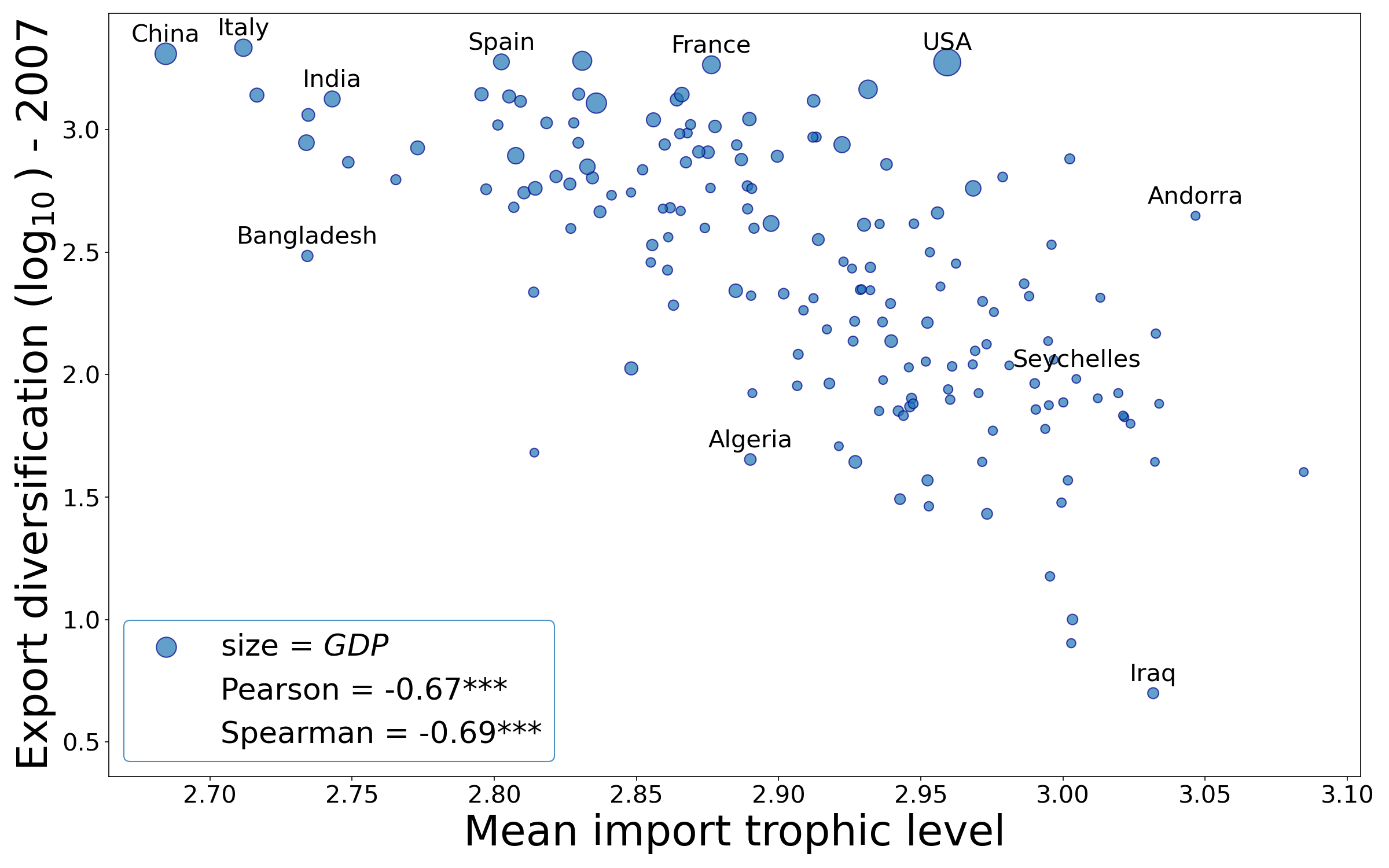}
    \end{subfigure}
    \begin{subfigure}{0.02\textwidth}
    \textbf{b)}
    \end{subfigure}
    \begin{subfigure}[t]{0.47\textwidth}  \includegraphics[width=\textwidth,valign=t]{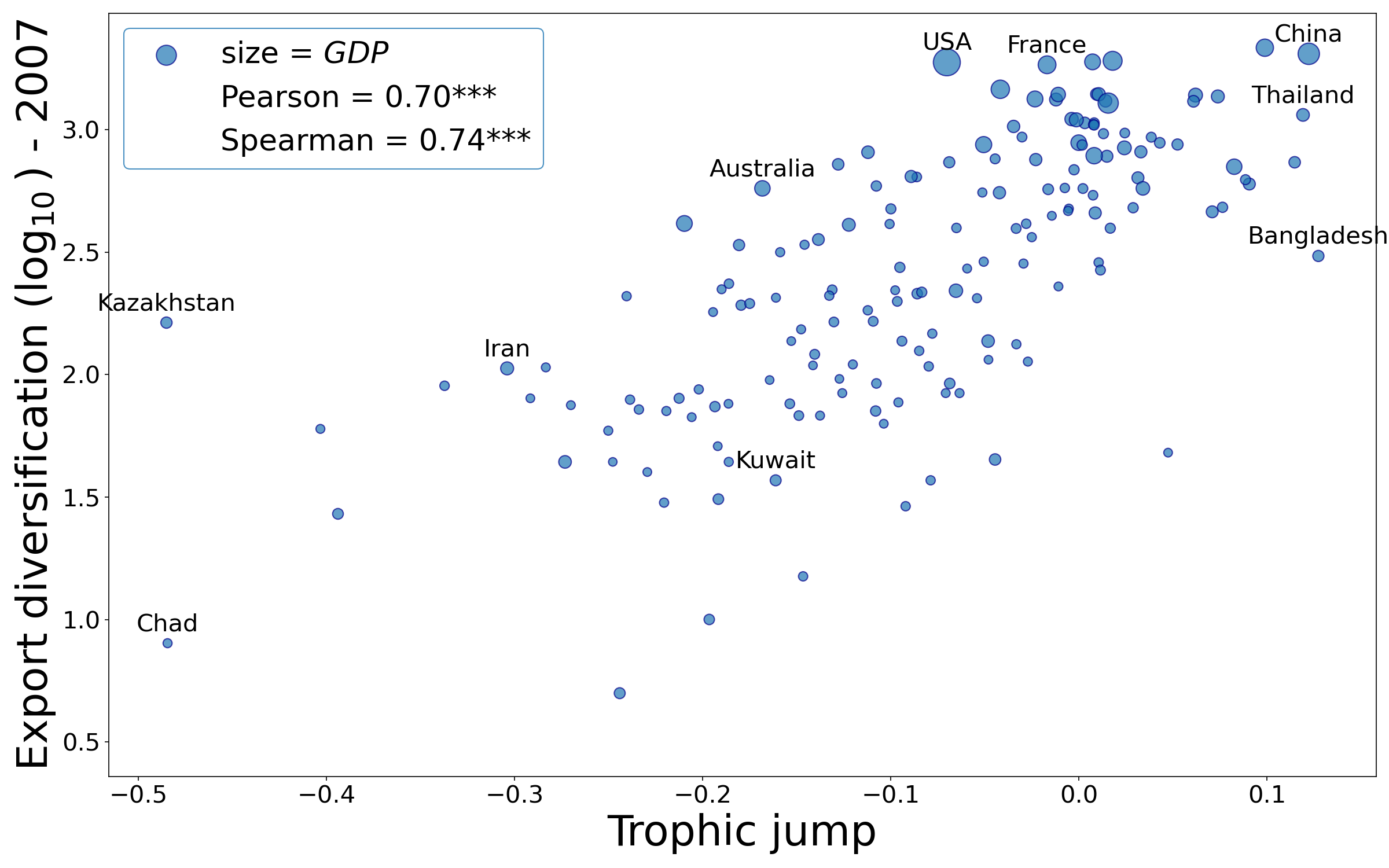}
    \end{subfigure}
    \begin{subfigure}{0.02\textwidth}
    \textbf{c)}
    \end{subfigure}
    \begin{subfigure}[t]{0.47\textwidth}  \includegraphics[width=\textwidth,valign=t]{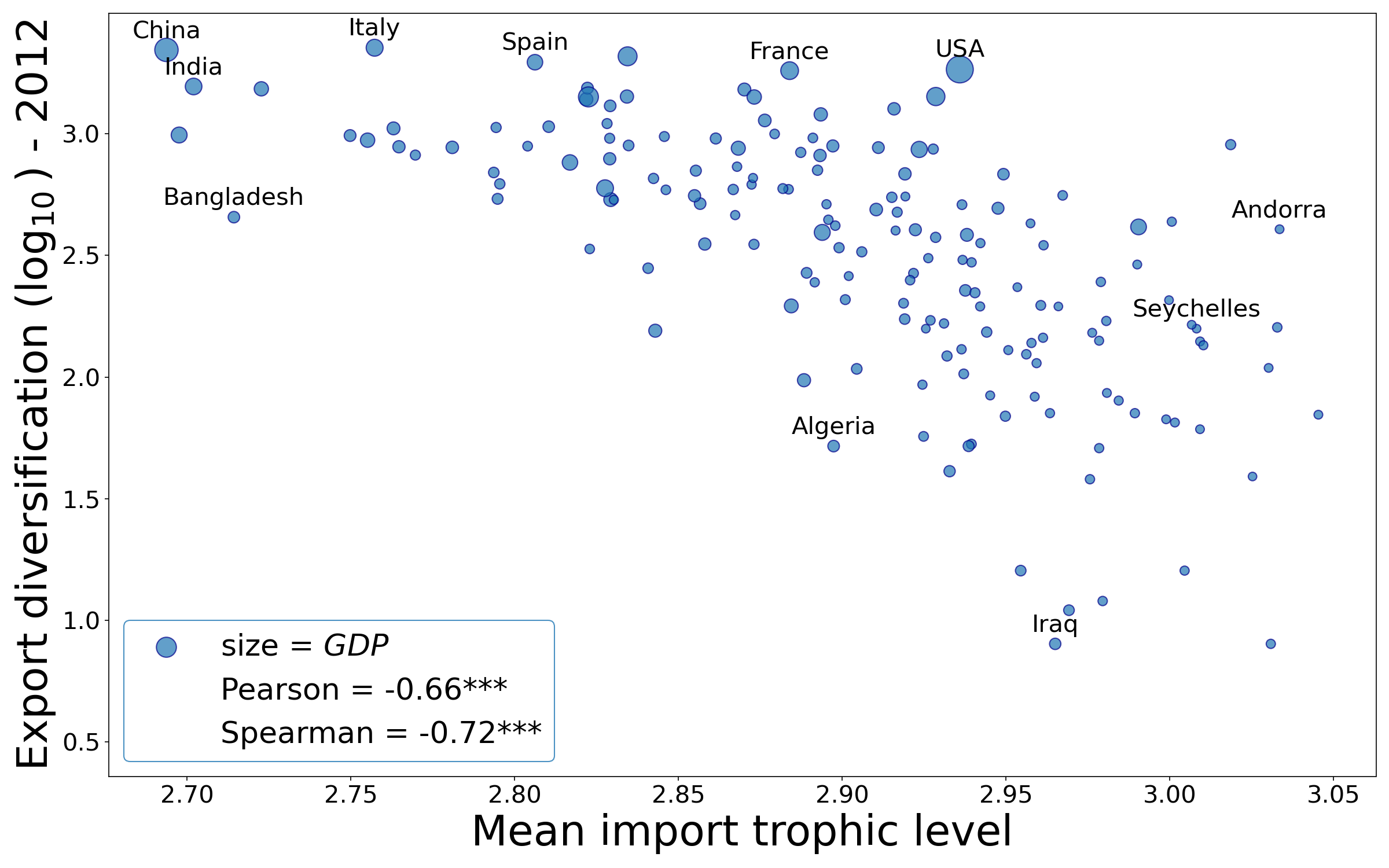}
    \end{subfigure}
    \begin{subfigure}{0.02\textwidth}
    \textbf{d)}
    \end{subfigure}
    \begin{subfigure}[t]{0.47\textwidth}  \includegraphics[width=\textwidth,valign=t]{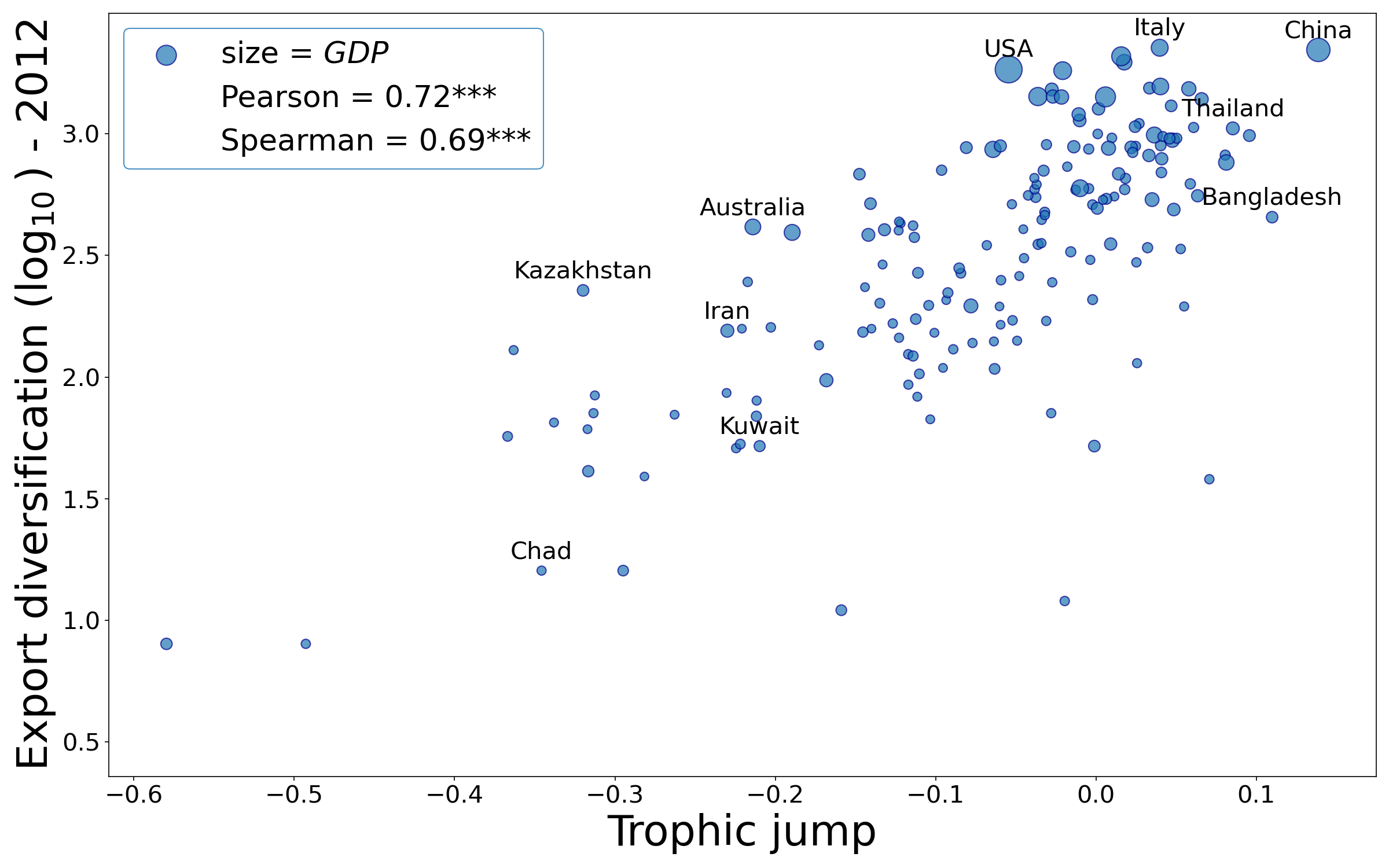}
    \end{subfigure}
    \begin{subfigure}{0.02\textwidth}
    \textbf{e)}
    \end{subfigure}
    \begin{subfigure}[t]{0.47\textwidth}  \includegraphics[width=\textwidth,valign=t]{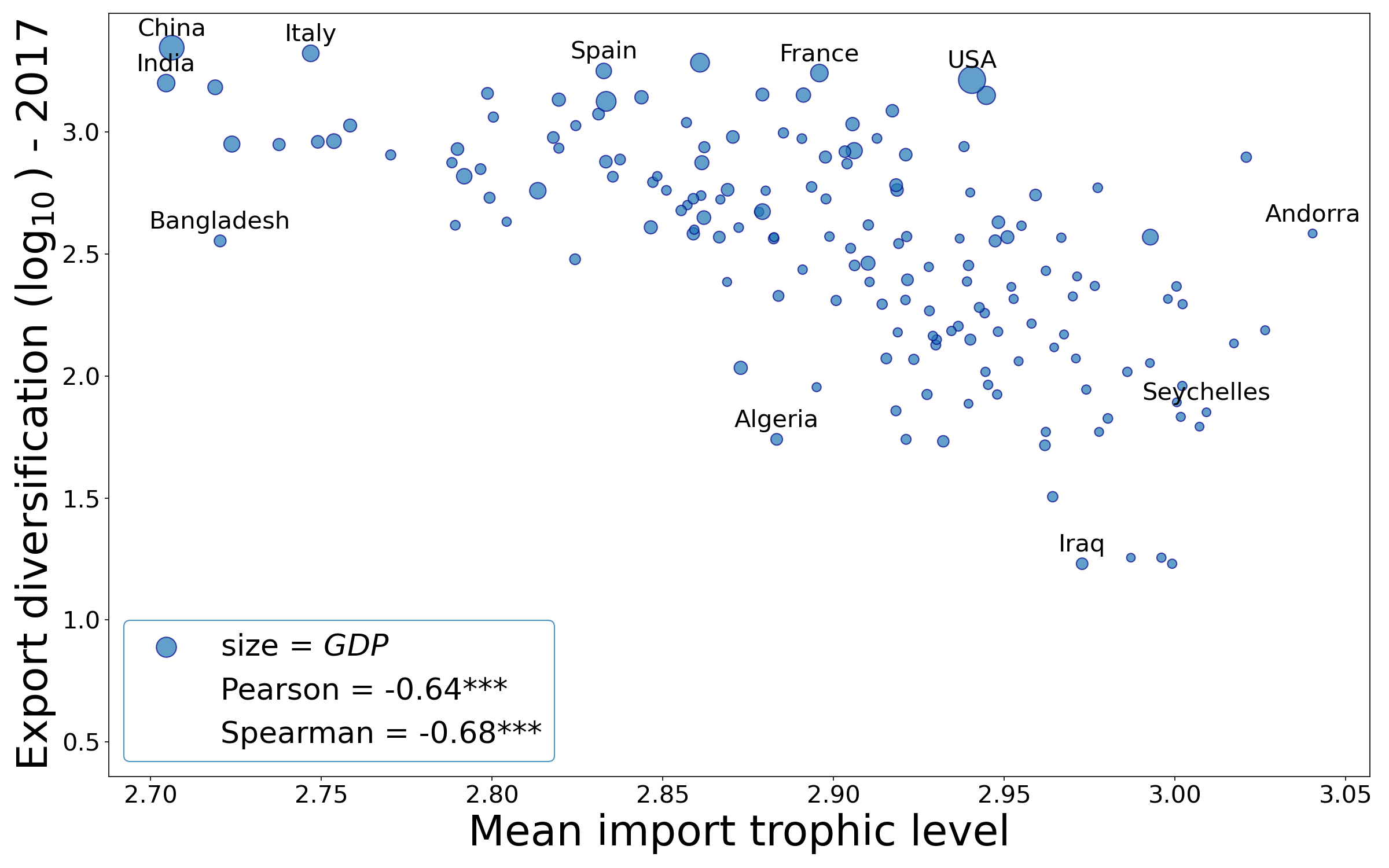}
    \end{subfigure}
    \begin{subfigure}{0.02\textwidth}
    \textbf{f)}
    \end{subfigure}
    \begin{subfigure}[t]{0.47\textwidth}  \includegraphics[width=\textwidth,valign=t]{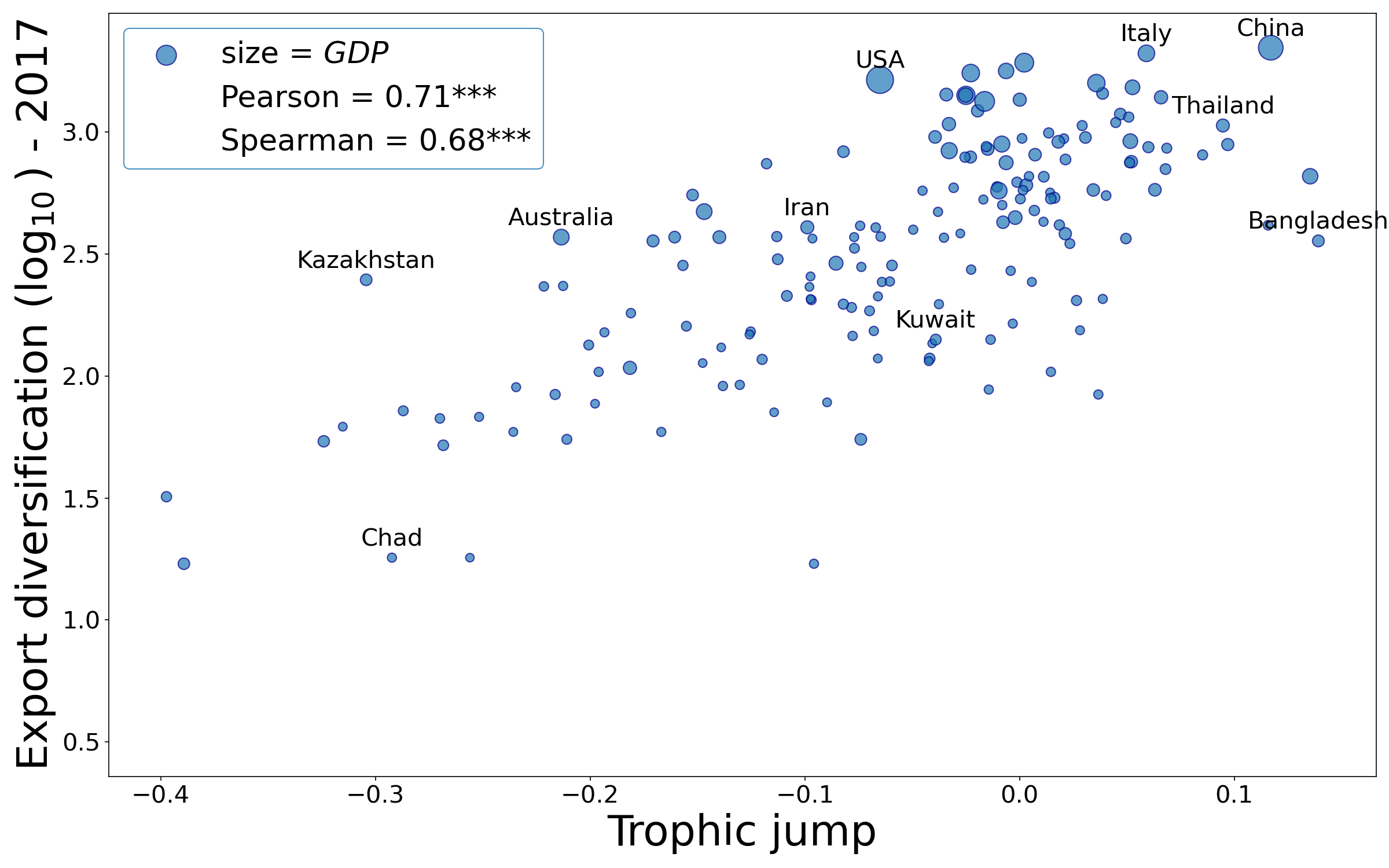}
    \end{subfigure}
    \caption{Temporal robustness of the observed trends for export diversification versus average import trophic level (panels \textbf{a}, \textbf{c} and \textbf{e}) and for export diversification versus trophic jump (panels \textbf{b}, \textbf{d} and \textbf{f}). The trends remain unvaried when looking at countries' imports and exports for 2007 (panels \textbf{a} and \textbf{b}), 2012 (panels \textbf{a} and \textbf{b}) and 2017 (panels \textbf{a} and \textbf{b}), exhibiting the same patterns obtained for 2023 (see figures 5 and 6 in the main text). Asterisks denote the significance level of the correlation coefficients: $p\leq0.01$ (***)}
    \label{fig:div_vs_tl_07_22}
\end{figure}

Regarding point a), in figure \ref{fig:div_vs_tl_07_22} we show the plots for export diversification versus average import trophic level and export diversification versus trophic jump for 2007 (panels \textbf{a} and \textbf{b}), 2012 (panels \textbf{c} and \textbf{d}) and 2017 (panels \textbf{e} and \textbf{f}): despite some changes in the positions of single countries, whose import and export basket variate along the years, the trends exhibited by both plots are consistent for all years (see also figures 5 and 6 in the main text).

Both trends are also observed if we replace the trophic levels computed on the products' network built from 2007 firms' import export data with their average over the time span 2007-2011, as shown in figure : for each year we independently build the products' network, compute the trophic levels (see Methods section in the main text), and then take the average for every product.

\begin{figure}[t!]
    \centering
    \begin{subfigure}{0.02\textwidth}
    \textbf{a)}
    \end{subfigure}
    \begin{subfigure}[t]{0.47\textwidth}  \includegraphics[width=\textwidth,valign=t]{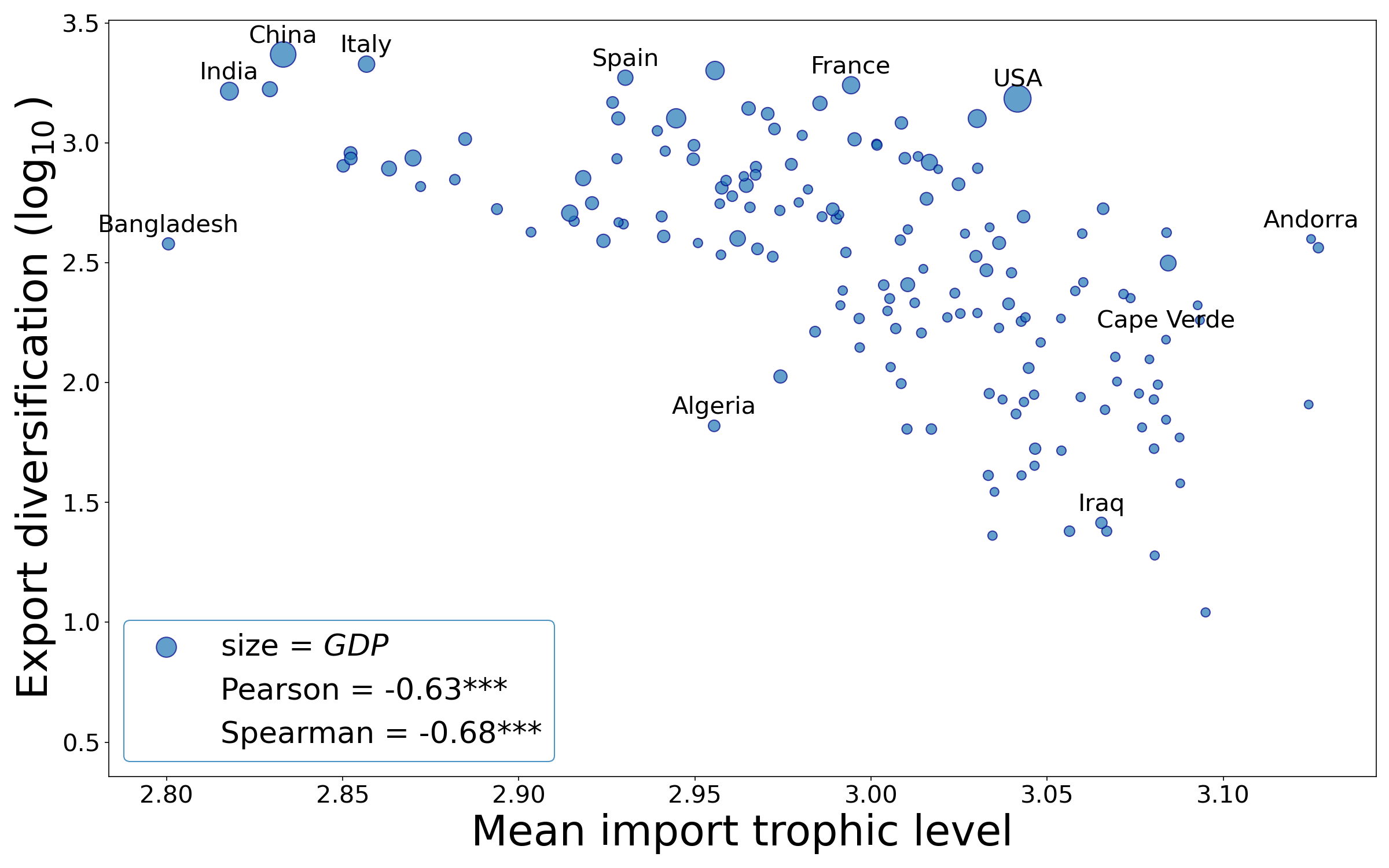}
    \end{subfigure}
    \begin{subfigure}{0.02\textwidth}
    \textbf{b)}
    \end{subfigure}
    \begin{subfigure}[t]{0.47\textwidth}  \includegraphics[width=\textwidth,valign=t]{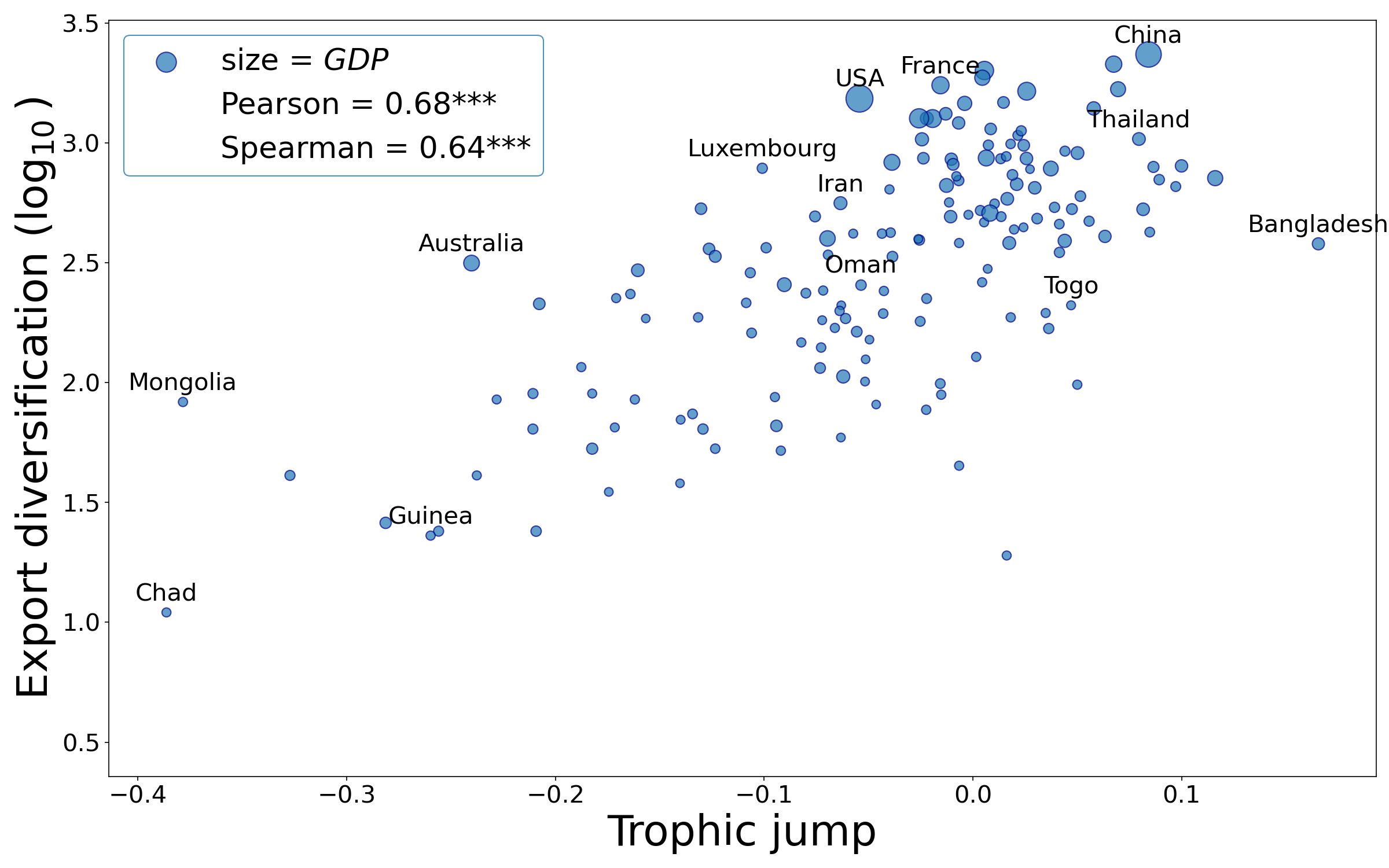}
    \end{subfigure}
    \caption{Export diversification versus average import trophic level (panel \textbf{a}) and export diversification versus trophic jump (panel \textbf{b}), with trophic levels averaged over the time span 2007-2011: despite slight changes in the correlation vaules, the trends are the same as the one observed using trophic levels for 2007 (see figures 5 and 6 in the main text). Asterisks denote the significance level of the correlation coefficients: $p\leq0.01$ (***)}
    \label{fig:div_vs_tl_07_11}
\end{figure}

\section{Growth regression for different time span}

We performed the linear regression on the logarithmic growth of $\text{GDP}_{pc}$ for $\Delta=5$ years: the results, reported in table \ref{tab:growth_reg_5}, confirm that, analogously to the case with $\Delta=5$ years, the model including both trophic level based variables has the strongest predictive power (i.e. highest value of adjusted $R^{2}$), and both variable are statistically significant. The same conclusions are drawn upon substituting the trophic levels computed from 2007 data with their average over the span 2007-2011, as shown in tables \ref{tab:growth_reg_5_07_11} - \ref{tab:growth_reg_10_07_11}.

\begin{table}[h!]
\centering
\begin{tabular}{lllllll}
\hline
 & Baseline & TL1 & TL2 & TL3 & EXP & ETL2\\
\hline
$GDP_{pc}(y)$ (log)       & -0.024*** & -0.035*** & -0.026*** & -0.038***         & -0.042***       & -0.045*** \\
Population (log)          &  0.028*** & 0.014     &  0.024*** &  0.013             & 0.006           & 0.004 \\
Trade \% of GDP (log)     &  0.035*** & 0.031***  &  0.033*** &  0.032***          & 0.027***        & 0.027*** \\
Constant                  &  0.079*** & 0.083***  &  0.081*** &  0.082***          & 0.081***        & 0.081*** \\
NITL                      &           & -0.036*** &           & \textbf{-0.056***} &                 & \\
NTJ                       &           &           & 0.015     & \textbf{-0.024**}  &                 & -0.013 \\
Export div. (log)         &           &           &           &                    &  0.042***       & 0.053*** \\
\hline
Adj. $R^{2}$              & 0.086     & 0.147     & 0.098     & \textbf{0.160}     & 0.148           & 0.153 \\
N° observations           & 1831      & 1831      & 1831      & 1831               & 1831            & 1831 \\
\end{tabular} 
\caption{Linear regression to predict $\log\left(\frac{\text{GDP}_{pc}(y+\Delta)}{\text{GDP}_{pc}(y)}\right)$, using a set of variables in year $y$, with $\Delta=5$ years. As in the case with $\Delta=10$ years (see table 2 in the main text), the model involving both trophic level-based covariates (TL3) displays the highest value of adjusted $R^{2}$, i.e. is the most predictive one. Asterisks denote different statistical significance levels of the variables: $p\leq0.1$ (*), $p\leq0.05$ (**), $p\leq0.01$ (***).}
\label{tab:growth_reg_5}
\end{table}

\begin{table}[h!]
\centering
\begin{tabular}{lllllll}
\hline
 & Baseline & TL1 & TL2 & TL3 & EXP & ETL2\\
\hline
$GDP_{pc}(y)$ (log)       & -0.024*** & -0.035*** & -0.026*** & -0.038***          & -0.042***  & -0.045*** \\
Population (log)          &  0.028*** & 0.014     &  0.024*** &  0.013             & 0.006      & 0.004     \\
Trade \% of GDP (log)     &  0.035*** & 0.031***  &  0.033*** &  0.031***          & 0.027***   & 0.027***  \\
Constant                  &  0.079*** & 0.083***  &  0.081*** &  0.081***          & 0.081***   & 0.081***  \\
NITL                      &           & -0.036*** &           & \textbf{-0.055***} &            &           \\
NTJ                       &           &           & 0.016*    & \textbf{-0.022*}   &            & -0.012    \\
Export div. (log)         &           &           &           &                    &  0.042***  & 0.052***  \\
\hline
Adj. $R^{2}$              & 0.086     & 0.147     & 0.101     & \textbf{0.158}     & 0.148      & 0.151 \\
N° observations           & 1831      & 1831      & 1831      & 1831               & 1831       & 1831 \\
\end{tabular} 
\caption{Linear regression to predict $\log\left(\frac{\text{GDP}_{pc}(y+\Delta)}{\text{GDP}_{pc}(y)}\right)$, using a set of variables in year $y$, with $\Delta=5$ years. In this case we replaced trophic levels from 2007 data with their average over the time span 2007-2011. Also in this case the model involving both trophic level based variables displays the highest value of adjusted $R^{2}$, i.e. is the most predictive one. Asterisks denote different statistical significance levels of the variables: $p\leq0.1$ (*), $p\leq0.05$ (**), $p\leq0.01$ (***).}
\label{tab:growth_reg_5_07_11}
\end{table}

\begin{table}[h!]
\centering
\begin{tabular}{lllllll}
\hline
 & Baseline & TL1 & TL2 & TL3 & EXP & ETL2\\
\hline
$GDP_{pc}(y)$ (log)       & -0.045*** & -0.071*** & -0.053*** & -0.075***          & -0.088***  & -0.093*** \\
Population (log)          &  0.050*** & 0.021     &  0.040**  &  0.017             & -0.001      & -0.005     \\
Trade \% of GDP (log)     &  0.055*** & 0.049***  &  0.051*** &  0.049***          & 0.041**    & 0.040**   \\
Constant                  &  0.157*** & 0.165***  &  0.163*** &  0.162***          & 0.163***   & 0.161***  \\
NITL                      &           & -0.078*** &           & \textbf{-0.115***} &            &           \\
NTJ                       &           &           & 0.039**   & \textbf{-0.042**}  &            & -0.018    \\
Export div. (log)         &           &           &           &                    &  0.096***  & 0.113***  \\
\hline
Adj. $R^{2}$              & 0.075     & 0.197     & 0.114     & \textbf{0.213}     & 0.202      & 0.205 \\
N° observations           & 1059      & 1059      & 1059      & 1059               & 1059       & 1059 \\
\end{tabular} 
\caption{Linear regression to predict $\log\left(\frac{\text{GDP}_{pc}(y+\Delta)}{\text{GDP}_{pc}(y)}\right)$, using a set of variables in year $y$, with $\Delta=10$ years. In this case we replaced trophic levels from 2007 data with their average over the time span 2007-2011. The model involving both trophic level based variables displays the highest value of adjusted $R^{2}$, i.e. is the most predictive one. Asterisks denote different statistical significance levels of the variables: $p\leq0.1$ (*), $p\leq0.05$ (**), $p\leq0.01$ (***).}
\label{tab:growth_reg_10_07_11}
\end{table}

\section{Growth regression with variables residualization}

As a robustness check for the regression results shown in table 2, we address collinearity between the normalized average import trophic level (NITL), the normalized trophic jump (NTJ) and the export diversification (EXP) by orthogonalizing the latter two with respect to the former, via residualization\cite{angrist2009mostly}. In detail, this means to perform the two linear regressions

\begin{equation}
    NTJ(y) = \alpha_{1} + \beta_{1}NITL(y) + \epsilon_{1}
\end{equation}

\begin{equation}
    EXP(y) = \alpha_{2} + \beta_{2}NITL(y) + \epsilon_{2}
\end{equation}

\noindent and take the residual errors $\epsilon_{1}$ and $\epsilon_{2}$ as substitutes of NTJ and EXP, respectively. The rationale is that such residuals represent the components of the original variables which are "\emph{not explained}" by NITL, i.e. the orthogonal components. In table \ref{tab:growth_reg_10_res} we show the results for the regression performed with this transformed variables, which confirm the findings reported in the main text: both NITL and NTJ are strong predictors of future $GDP_{pc}$ growth, with negative, significant coefficients. The residual of EXP, on the other hand, is never significant, meaning that, once the normalized average trophic level is specified, the export diversification does not add information.

\begin{table}[h!]
\centering
\begin{tabular}{llllll}
\hline
 & Baseline & TL1 & TL2R & EXPR & ETL2R\\
\hline
$\text{GDP}_{pc}$ (log)      & -0,045*** & -0,071*** & -0.075***          & -0,083*** & -0,085***          \\
Population (log)             & 0,050***  &  0,020    & 0.018              &  0,006    & 0.006              \\
Trade \% of GDP (log)        & 0,057***  &  0,048*** & 0.050***           &  0,043*** & 0,046***           \\
Constant                     & 0,155***  &  0,165*** & 0.163***           &  0,164*** & 0,163***           \\
NITL                         &           & -0,078*** & \textbf{-0,081***} & -0.088*** & \textbf{-0,089***} \\
NTJ                          &           &           & \textbf{-0,028**}  &           & \textbf{-0,026**}  \\
Exp div (log)                &           &           &                    &  0.022    &  0.017             \\
\hline
Adj. $R^{2}$                 & 0,075     & 0,198     & \textbf{0,218}     & 0.205     & \textbf{0,222}     \\
N° observations              & 1.059     & 1.059     & 1.059              & 1.059     & 1.059              \\
\end{tabular}
\caption{Linear regression to predict $\log\left(\frac{\text{GDP}_{pc}(y+\Delta)}{\text{GDP}_{pc}(y)}\right)$, using a set of variables in year $y$, with $\Delta=10$ years. In order to break the collinearity between NITL, NTJ and export diversification, the latter have been regressed on the former, and the corresponding residuals have been kept as variables. The results confirm NITL and NTJ to be strong predictors of future growth, as the models including both (TL2R and ETL2R) display the highest predictive power. The residual of export diversification, on the other hand, is not significant, and its inclusion results just in a slight increase of adjusted $R^{2}$. Asterisks denote different statistical significance levels of the variables: $p\leq0.1$ (*), $p\leq0.05$ (**), $p\leq0.01$ (***).}
\label{tab:growth_reg_10_res}
\end{table}


\end{document}